\documentclass[11pt,letterpaper,final]{article}

\usepackage{amsmath}
\usepackage{mathtools}
\usepackage{amsfonts}
\usepackage{amssymb}
\usepackage{graphicx}
\usepackage{color}
\usepackage[left=1in,top=1in,right=1in,bottom=1in]{geometry}
\usepackage{multirow}
\usepackage{float}
\usepackage{array}
\usepackage{svg}
\usepackage{lineno}
\usepackage{setspace}
\usepackage[numbers,sort&compress]{natbib}
\usepackage{bm}
\usepackage{physics}
\usepackage{authblk}
\usepackage{caption}
\usepackage{subcaption}
\usepackage{hyperref}
\hypersetup{
colorlinks=true,
linkcolor=black,
filecolor=black,      
urlcolor=black,
citecolor=black,
}
\usepackage{mathrsfs}
\usepackage[version=4]{mhchem}

\usepackage[utf8]{inputenc}
\usepackage[T1]{fontenc}
\usepackage{lmodern}
\usepackage{chemformula}

\begin{document}

\title{Biomimetic non-ergodic aging by dynamic-to-covalent transitions in physical hydrogels}

\author[1]{Samya~Sen}
\author[1]{Anthony~C.~Yu}
\author[1]{Changxin~Dong}
\author[1]{Andrea~I.~D'Aquino}
\author[1,2,3,4,5,6]{Eric~A.~Appel\thanks{Corresponding author: eappel@stanford.edu}}

\affil[1]{Department of Materials Science \& Engineering, Stanford University, Stanford, CA 94305, USA}
\affil[2]{Department of Bioengineering, Stanford University, Stanford, CA 94305, USA}
\affil[3]{Stanford ChEM-H, Stanford University, Stanford, CA 94305, USA}
\affil[4]{Institute for Immunity, Transplantation and Infection, Stanford University School of Medicine, 
Stanford, CA 94305, USA}
\affil[5]{Department of Pediatrics - Endocrinology, Stanford University School of Medicine, Stanford, 
CA 94305, USA}
\affil[6]{Woods Institute for the Environment, Stanford University, Stanford CA 94305, USA}
\setcounter{Maxaffil}{0}
\renewcommand\Affilfont{\itshape\large}




\date{}

\maketitle

\setstretch{1.5}

\begin{abstract}

Hydrogels are soft materials engineered to suit a multitude of applications that exploit their tunable mechanochemical properties. Dynamic hydrogels employing noncovalent, physically crosslinked networks dominated by either enthalpic or entropic interactions enable unique rheological and stimuli-responsive characteristics. In contrast to enthalpy-driven interactions that soften with increasing temperature, entropic interactions result in largely temperature-independent mechanical properties. By engineering interfacial polymer-particle interactions, we can induce a dynamic-to-covalent transition in entropic hydrogels that leads to biomimetic non-ergodic aging in the microstructure without altering the network mesh size. This transition is tuned by varying temperature and formulation conditions such as $p$H, which allows for multivalent tunability in properties. These hydrogels can thus be designed to exhibit either temperature-independent metastable dynamic crosslinking or time-dependent stiffening based on formulation and storage conditions, all while maintaining strucutural features critical for controlling mass transport, akin to many biological tissues. Such robust materials with versatile and adaptable properties can be utilized in applications such as wildfire suppression, surgical adhesives, and depot-forming injectable drug delivery systems.

\end{abstract}


\section{\label{sec:intro}Introduction}
Physically crosslinked soft materials exhibit many unique functional and stimuli-responsive physical and mechanical properties \cite{Appel_CSR2012,Mann2017,Webber2016,Seiffert2015}. Over the years, soft materials employing either enthalpy- or entropy-driven dynamic noncovalent crosslinking interactions have been designed to have unique mechano\-chemical properties tunable to a fine degree based on target behaviors \cite{Anthony2021}. These types of dynamic network materials, including many examples of physical hydrogels, have been employed in a variety of applications that exploit their robust and versatile properties, including biomaterials \cite{Rodell2013,Rodell2016,Hernandez2018,Grosskopf2019,Wong2009}, drug delivery \cite{Buwalda2017,Li2016,Mitragotri2014,Mann2017,Webber2016}, self-healing materials \cite{Ma2021,Clarke2017,Rao2016,Cordier2008,Li2016_SH}, wearable electronics \cite{Rao2016,Oh2016,Yan2018,Li2016_SH,Someya2016}, 3D printing \cite{Ouyang2016,Dubbin2016,Murphy2014}, and sprayable coatings for wildland fire suppression \cite{AnthonyPNAS2016,AnthonyPNAS2019}.

The dynamic physical hydrogel literature is rife with materials that leverage enthalpy-dominated crosslinking interactions, which are inherently thermoresponsive \emph{via} temperature-dependent viscoelastic properties. Such materials generally soften at elevated temperatures owing to the highly specific enthalpy-driven binding reactions including ionic, host-guest, hydrogen bonding, and hydrophobic interactions \cite{Mann2017,Rodell2013,Clarke2017,Rao2016,Oh2016,Yan2018,Cordier2008,Webber2016,Appel2010}. More recently,  entropy-driven hydrogels with dynamic networks based on multivalent polymer-particle interactions utilizing carefully tuned interfacial chemistry have been formulated \cite{AnthonyPNAS2016,Anthony2021,Anthony2019,Hernandez2018}. Such networks built with entropy-dominated crosslinking interactions can exhibit temperature-independent viscoelasticity, and this unique property has been leveraged in diverse applications from fire suppression \cite{AnthonyPNAS2016} to drug delivery \cite{Correa2021} where consistent mechanical properties despite temperature variations are necessary. 

\begin{figure}[!ht]
\centering
\includegraphics[width=\textwidth]{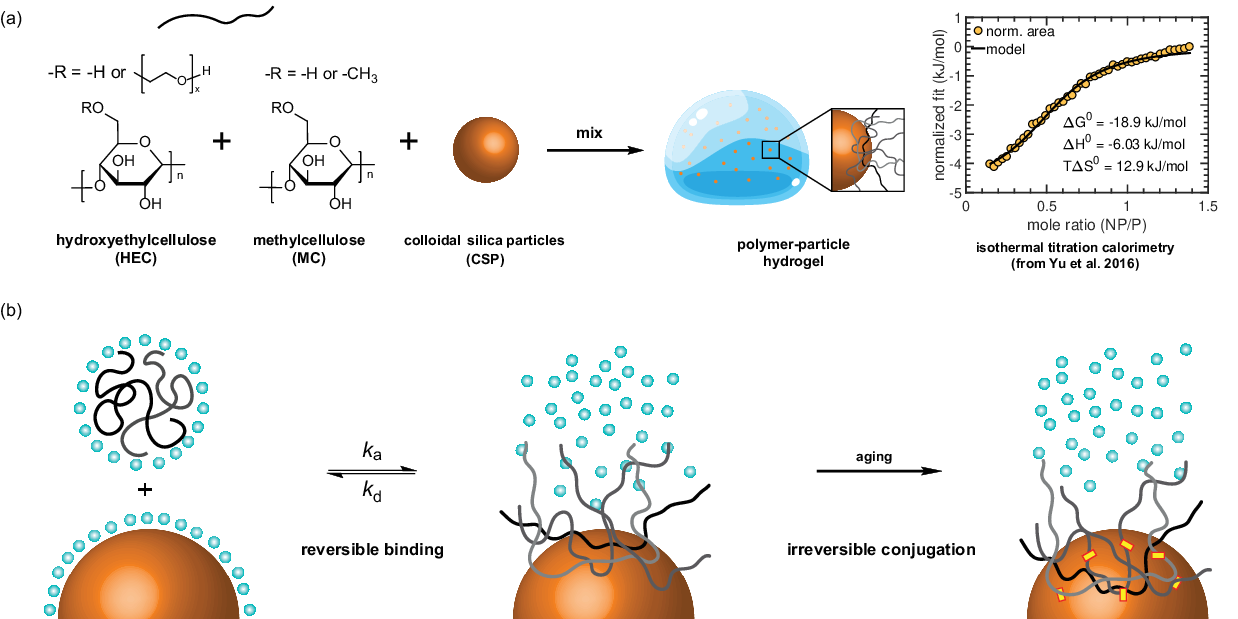}
\caption{Schematic representation of the chemical aging and proliferation of crosslinks in the hydrogels with age. (a) HEC and MC polymer chains when mixed with CSP dispersion form a hydrogel, where the CSPs function as transient crosslink junctions between polymer chains adsorbed onto the surface. The plot shows the ITC data from Yu \emph{et al.} \cite{AnthonyPNAS2016} for calculating the free energy of binding $\Delta G^0$, with the enthalpic, $\Delta H^0$, and entropic, $T\Delta S^0$, contributions, for the entropy-dominated hydrogels. (b) Progression of interactions between polymer chains and particle surface studied in this work, with dynamic, entropy-driven reversible binding at short times after mixing, followed by the formation of irreversible enthalpy-driven covalent bonds between the CSP and the adsorbed HEC-MC chains due to aging.}
\label{fig:schematic-aging}
\end{figure}

In this work, we evaluate an example entropy-driven hydrogel system where mixing of solutions of cellulose derivatives (HEC and MC) with colloidal silica particles (CSP) generate dynamic HEC-MC/CSP hydrogels with facile and scalable formulation (Fig.~\ref{fig:schematic-aging}a)
This material platform exhibits viscoelastic rheological properties largely independent of temperature, attributed to the dominance of the entropic contribution to free energy of the polymer-particle interactions. We report a curious effect in this system where when aged, the material gets stiffer over time (i.e., the modulus increases with aging time) without alteration of the hydrogel microstructure. This aging behavior is observed consistently over the course of weeks, and is dependent on storage temperature and formulation pH, suggesting activated processes drive a transition from dynamic to covalent crosslinking interactions in the system. Such behaviors are fundamentally akin to aging observed in the extracellular matrix found in many biological tissues and enable advantageous multifaceted tunability of hydrogel mechanical properties by exploiting the differential temperature and $p$H dependence. The biomimetic dynamic-to-covalent transitions in these synthetic physical hydrogels enable the development of materials with structure defined by the self-assembly of the gel components that then evolve into covalent interactions in defined ways, making them highly distinguishing in numerous application spaces.

\section{\label{sec:exp}Experimental}

\subsection{\label{subsec:exo-matl}Materials}
The hydrogels studied in this work are prepared following a protocol outlined in the literature \cite{AnthonyPNAS2016}. First, polymer solutions were prepared by dissolving hydroxyethylcellulose or HEC ($M_{\rm v} \simeq 1,300$~kDa, Sigma) and methylcellulose or MC ($M_{\rm v} \simeq 90$~kDa, Sigma) in water (or citrate buffer for $p$H 4 gels) at a concentration of 3\%~w/w, with stirring and mild heating. Colloidal silica particles or CSP dispersions were made at 15\%~w/w by diluting Ludox TMA 34\%~w/w dispersions (acid stable, $D_{\rm H} \simeq 22$~nm) in water (or citrate buffer for $p$H 4 gels). Finally, equal weights of HEC-MC solutions and CSP dispersions were mixed well by vortexing to obtain 1.5-7.5\% w/w HEC-MC/CSP gels. The properties of the gels are independent of the scale of mixing as has already been shown in earlier studies \cite{AnthonyPNAS2016,AnthonyPNAS2019}.

\subsection{\label{subsec:exp-rheology}Rheometry}
Rheological characterization of the gels was performed on a separated motor-transducer rheometer (ARES-G2, TA Instruments) at Stanford Soft and Hybrid Materials Facility (SMF). Stress relaxation tests were done by applying a step strain $\gamma_0 = 2\%$ to materials and the evolution of the shear viscoelastic modulus $G(t; t_{\rm age}, T)$ was monitored over 1000~s, where $t_{\rm age}$ is the age of the sample and $T$ is the absolute temperature at which the gels were stored. For rheometry, the samples were loaded between 25~mm hatched parallel plates to mitigate wall slip artifacts, and the temperature was controlled by an Advanced Peltier System. Five sample ages were used for the aging study: 0 ($\sim1~{\rm h}$), 1, 3, 7, and 15 days. Three temperatures of storage were used to study the effect of temperature on the aging: 4, 25, 37$^\circ$C (at $p$H 7). Effect of $p$H was tested by making gels in sodium citrate buffer with $p$H 4 compared with gels made in water, stored at 25$^\circ$C. All tests were duplicated.

\subsection{\label{subsec:exp-ftir}FTIR spectroscopy}
The extent of reaction was inferred using Fourier Transform Infrared Spectroscopy (FTIR) on each sample. Attenuated Transmission Reflectance (ATR) setup was used to probe each sample on a Nicolet-iS50 spectrometer at  Stanford Soft and Hybrid Materials Facility (SMF). Each sample was aged for the requisite duration and lyophilized at the end of the aging process; the lyophilized powder was loaded on the diamond window of the ATR setup and the reflectance was observed using a \ch{KBr} beamsplitter for wavenumbers $\bar{\nu} = 400$-5000~cm$^{-1}$. Sample data was collected in duplicates. The absorbance spectrum of the material in this range of wavenumbers was used to further quantify the decay or proliferation of bonds due to aging.


\section{\label{sec:aging}Results}
In this work, we first discuss the rheological signatures of aging in our HEC-MC/CSP hydrogel system and comment on the thermodynamic nature of the aging process, along with the temperature and $p$H dependence. We then provide structural insights into this aging process obtained from diffusivity measurements, which suggests negligible alterations to the network structure during aging. Based on this, we describe the physics of the aging process and hypothesize the possible interfacial interactions that may underlie this unique behavior. We then provide corroborating evidence for the hypothesized chemical changes using infrared spectroscopy data that sheds more light on the nature of interfacial chemistry between the polymer-particle components of the hydrogel system, and hence rationalize the evolution of chemical bonds in the material with aging time. Following this, we model the chemical reaction underlying the aging process. We use second-order reaction kinetics to model the dynamic-to-covalent transition and infer microstructural changes in the material by connecting these kinetics to bulk mechanical properties using network theory. Finally, we summarize the findings of this paper and comment on the outlook for engineering soft materials with such novel and highly tunable mechanochemical properties.

It is common practice to study the stiffness of aging systems using mechanical perturbation, such as rheometry. Often, percolating systems, aggregating dispersions, soft glasses, and crowded suspensions are probed using stress relaxation, which provides long-time information about material properties while keeping the sample deformation within the linear regime \cite{Sollich1997,Sollich1998,Cloitre2000_PRL,Fielding2000,Baumberger2009,Petekidis2018,Fielding2020,SenJoR2022,Sen_PhDThesis,WangJoR2022}. In this work, we first determined the stiffness of our HEC-MC/CSP gels from the short-time plateau modulus for each gel formulation, $G_0$, as a function of aging time to quantify the aging process. We then studied the effect of temperature and $p$H on the aging process by monitoring the evolution of plateau modulus values over time. The plateau modulus can be connected to the microstructure of the gels using network theory, thus quantifying the various chemical reactions and structural evolution networks undergo when aged ({\em vide infra}).


\subsection{\label{subsec:aging-rheology-temp}Temperature-dependent aging}
\begin{figure}[!ht]
\centering
\begin{subfigure}[b]{0.32\textwidth}
	\centering
	(a)
	\includegraphics[width=\textwidth]{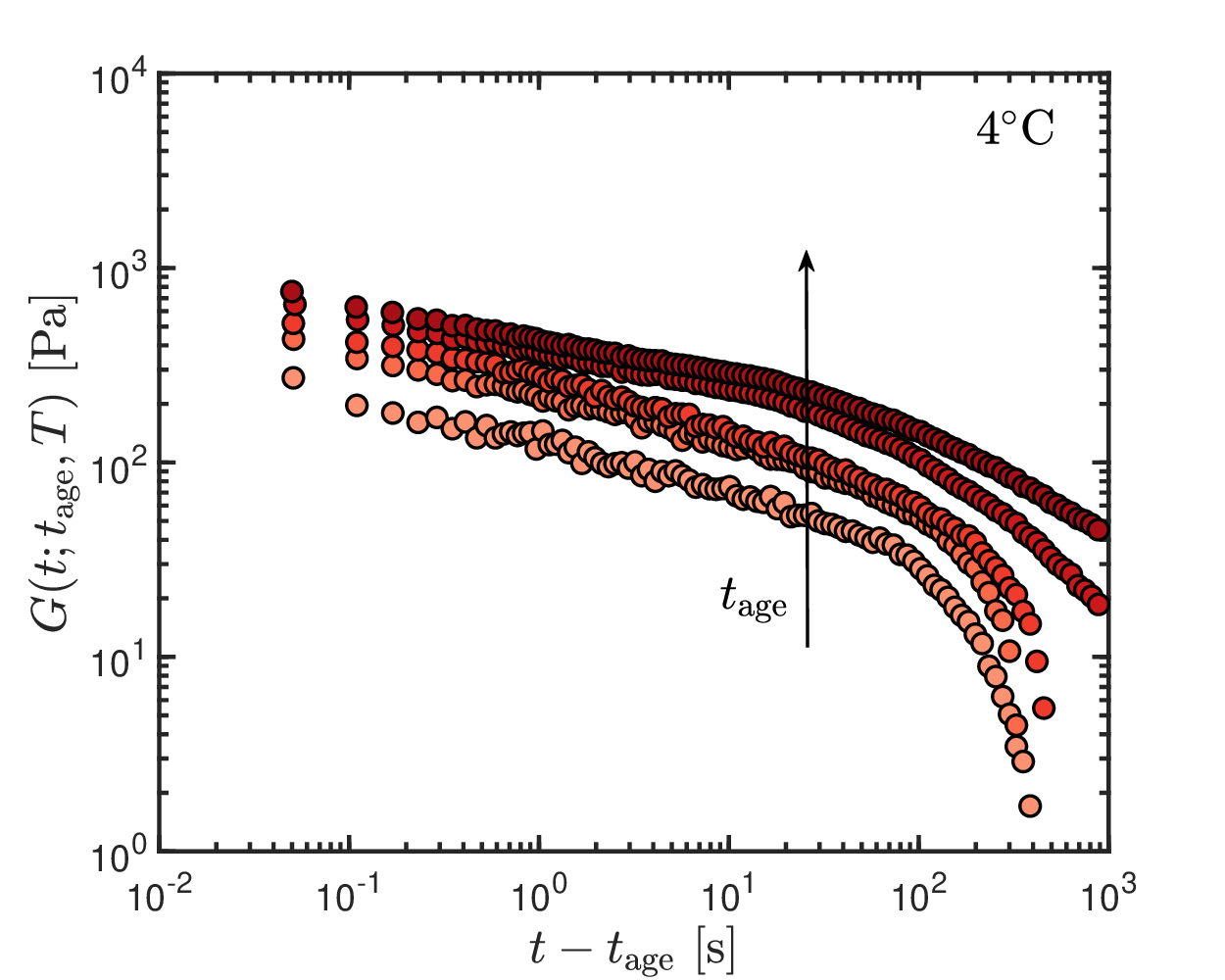}
\end{subfigure}
\begin{subfigure}[b]{0.32\textwidth}
	\centering
	(b)
	\includegraphics[width=\textwidth]{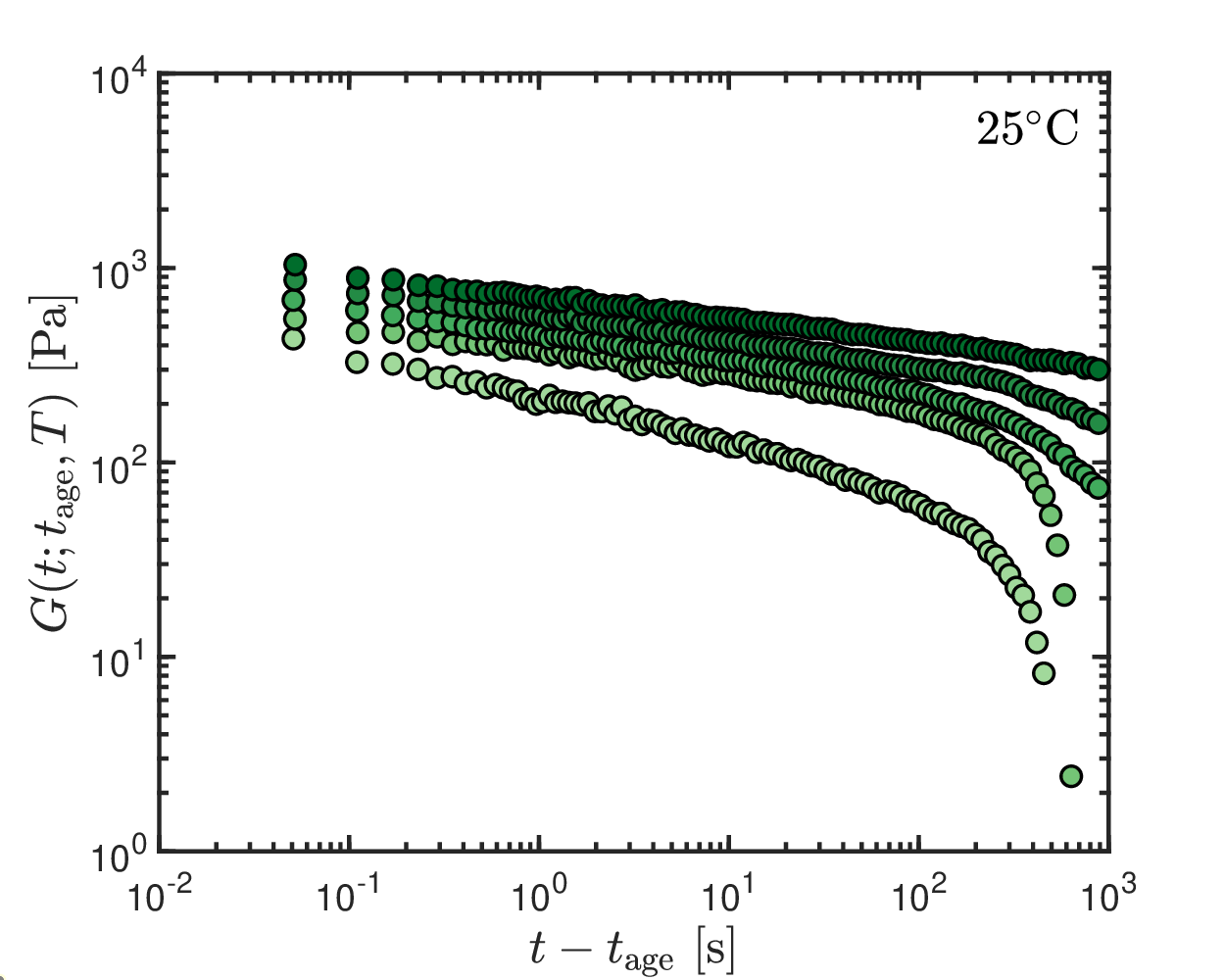}
\end{subfigure}
\begin{subfigure}[b]{0.32\textwidth}
	\centering
	(c)
	\includegraphics[width=\textwidth]{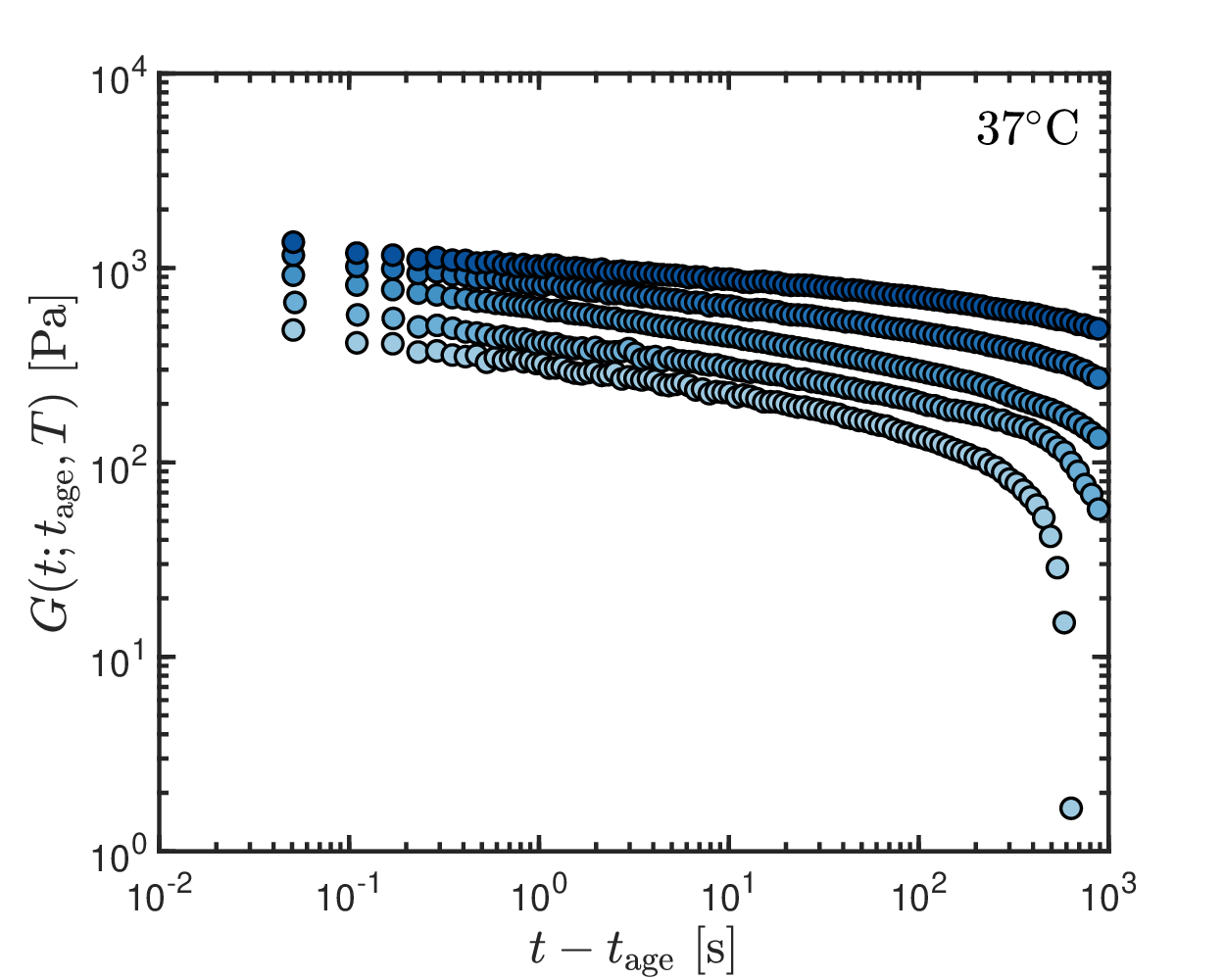}
\end{subfigure}
\caption{Stress relaxation for gels at different aging times at (a) 4$^\circ$C, (b) 25$^\circ$C, and (c) 37$^\circ$C. For a given temperature, gels get stiffer with age, and its relaxation timescale becomes longer. The same features are observed as temperature increases for a given age of the gel.}
\label{fig:G-relaxation-temp-time}
\end{figure}

Stress relaxation data for gels at different timepoints following aging at three different temperatures demonstrates that, for a given temperature, the gels stiffen with age (Fig.~\ref{fig:G-relaxation-temp-time}). As $t_{\rm age}$ increases, the relaxation curve shifts towards higher modulus, resulting in higher values of $G_0$. Additionally, the relaxation behavior of the gels also changes. The stress relaxation transients become broader and the characteristic timescale of relaxation becomes longer, which is evident from the slower decay of relaxation modulus with age. This observation suggests that the material possesses multiple timescales of relaxation, or a relaxation spectrum, and not just one single characteristic timescale \cite{DPL_vol1,Ferry}. Polymeric materials, especially those with transient crosslinks, often exhibit such behavior. These signatures of the stress relaxation behavior indicate a broadening in the dynamics of the gels with age, linked with the stiffening behavior and proliferation of stronger crosslinks through the material.

As the temperature increases, the aging becomes more pronounced. Indeed, the modulus increases faster with age, and the relaxation behavior becomes even broader. Clearly, temperature plays a key role in the structural change of the gels, and the reaction responsible for such a change proceeds faster at higher temperatures. These behaviors are characteristic of endothermic covalent reaction kinetics, which we hypothesized may occur over time between the alcohol groups on the cellulose chains and the silanol groups on the CSP surface, which are in close proximity to each other due to the strong yet dynamic non-covalent polymer-particle interactions. As such a condensation reaction would be expected to be $p$H-dependent, we then evaluated the effect of $p$H on aging.


\subsection{\label{subsec:aging-rheology-pH} \emph{p}H-dependent aging}
\begin{figure}[!ht]
\centering
\begin{subfigure}[b]{0.4\textwidth}
	\centering
	(a)
	\includegraphics[width=\textwidth]{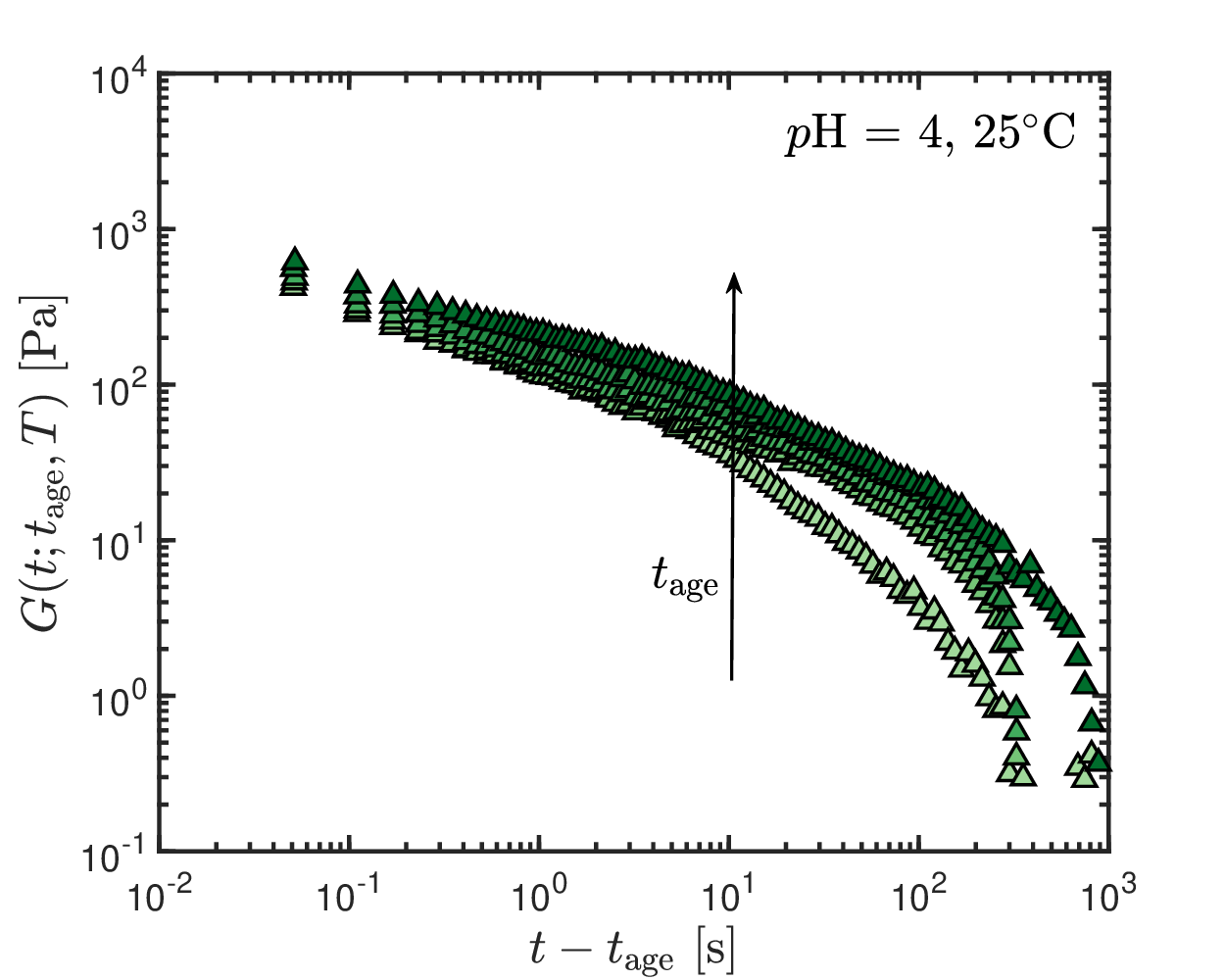}
\end{subfigure}
\begin{subfigure}[b]{0.4\textwidth}
	\centering
	(b)
	\includegraphics[width=\textwidth]{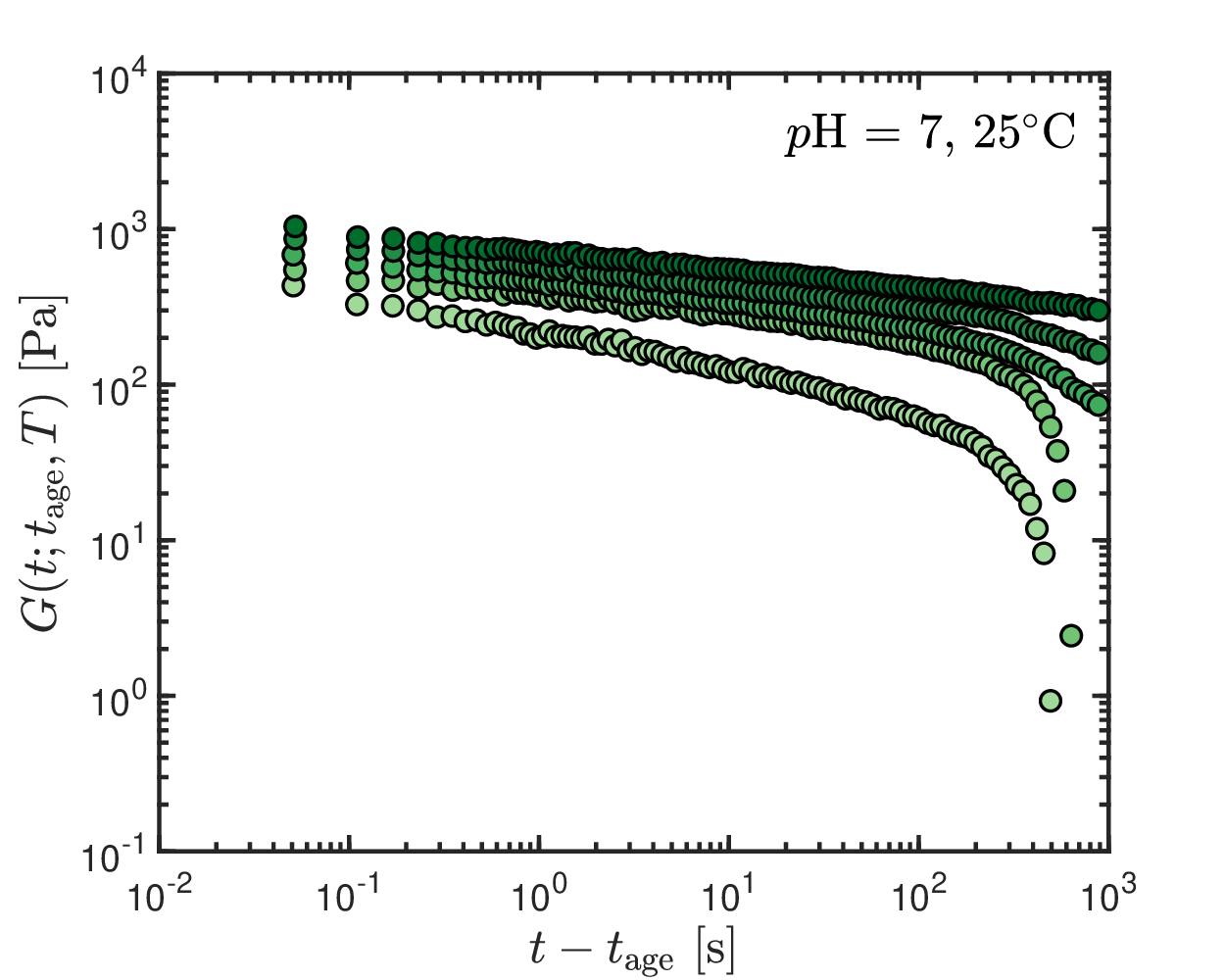}
\end{subfigure}
\caption{Stress relaxation for gels at different aging times at (a) $p{\rm H} = 4$, and (b) $p{\rm H} = 7$, all gels at $25^\circ$C. For a given $p$H, gels get stiffer with age, and its relaxation timescale becomes longer. The same features are observed as $p$H increases for a given age of the gel.}
\label{fig:G-relaxation-pH-time}
\end{figure}

Stress relaxation data for gels prepared at two $p$H values ($p{\rm H} \simeq 4$ and 7) and stored at 25$^\circ$C were obtained for different aging times (Fig.~\ref{fig:G-relaxation-pH-time}). Similar to our previous observations, the modulus increases and the relaxation spectrum gets broader with age. However, both the relative magnitudes of $G_0$ and the broadening of the relaxation spectrum is decreased at lower $p$H.

These data indicate that the reaction responsible for aging in the gels proceeds slower in more acidic conditions, while increasing temperature speeds up the reaction. These observations suggest that a potential driver of the aging process is an interfacial covalent reaction between the silanol groups on the CSPs and the alcohol groups on the cellulose polymers. Such a reaction would be expected to proceed faster at higher temperatures (whereby the enthalpic contribution to the free energy of change dominates), and faster in a more alkaline environment as deprotonated silanol groups are more reactive, favoring a condensation reaction with the cellulose alcohol groups. In the following section, we discuss the thermodynamic nature of these aging processes.


\subsection{\label{subsec:aging-rheology-ergodicity}Non-ergodicity in aging gels}	
\begin{figure}[!ht]
\centering
\begin{subfigure}[b]{0.32\textwidth}
	\centering
	(a)
	\includegraphics[width=\textwidth]{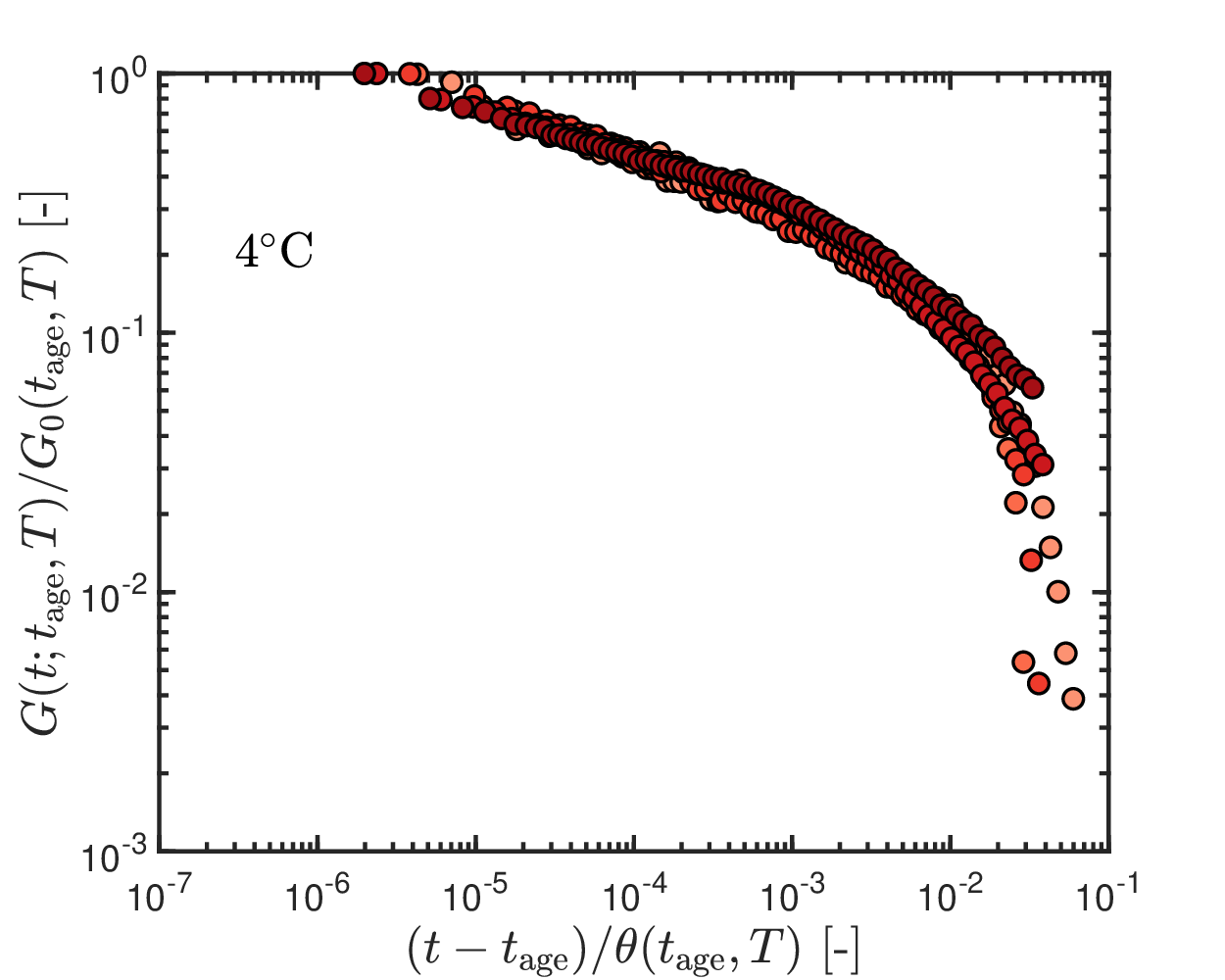}
\end{subfigure}
\begin{subfigure}[b]{0.32\textwidth}
	\centering
	(b)
	\includegraphics[width=\textwidth]{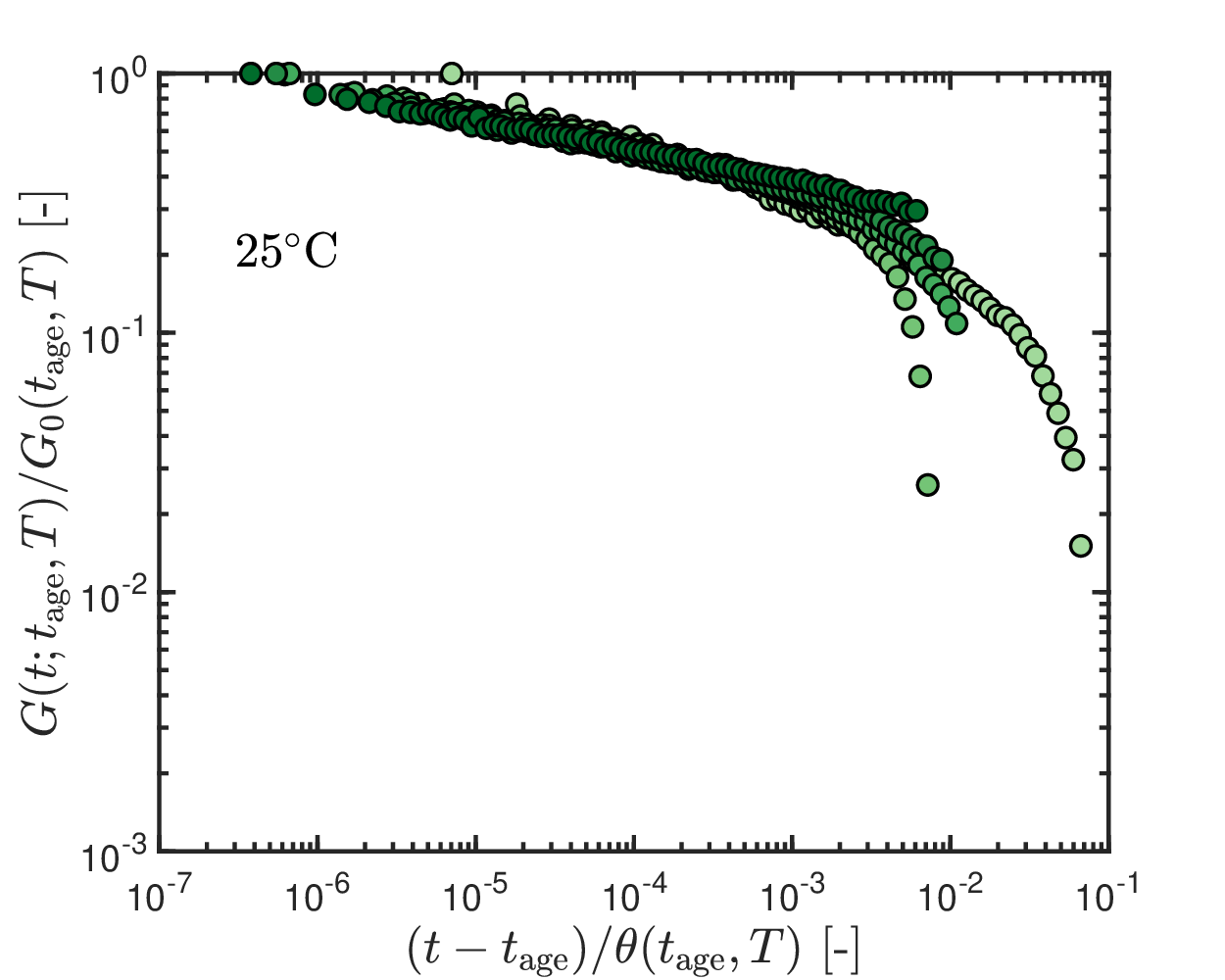}
\end{subfigure}
\begin{subfigure}[b]{0.32\textwidth}
	\centering
	(c)
	\includegraphics[width=\textwidth]{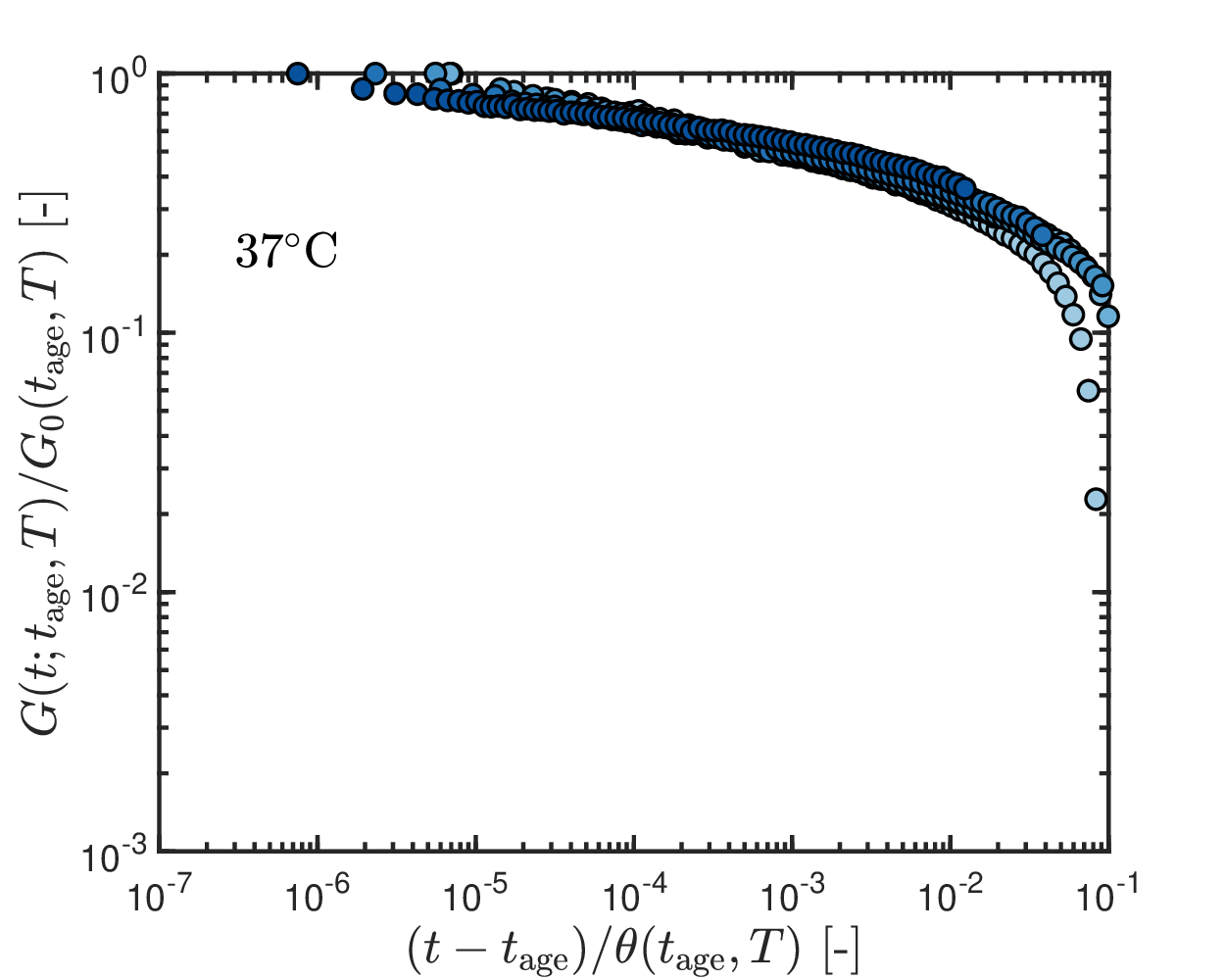}
\end{subfigure}
\caption{Master curves of stress relaxation data shifted to account for aging, for gels at different aging times at (a) 4$^\circ$C, (b) 25$^\circ$C, and (c) 37$^\circ$C. Such shifting captures non-ergodic effects in the material.}
\label{fig:G-relaxation-temp-time-shifted}
\end{figure}

Typically, for aging gel systems, the underlying process of aging is classified as non-ergodic (i.e., in violation of time-invariant physics) \cite{Fielding2000}. Briefly, non-ergodic physical processes proceed such that upon changing the starting point (i.e., equivalent to shifting the location of $t=0$), one changes the physics of the problem, thus breaking time-translation symmetry. The evolution of the relaxation spectrum of our HEC-MC/CSP gels with age corroborates this observation. The non-ergodicity is captured using a nonlinear time-axis shift that, in theory, should collapse the data. The shifting is based on the time of aging itself, $t_{\rm age}$, and a shifting exponent, $\mu$, which generates master curves for the aging gels \cite{Cloitre2000_PRL,Fielding2000,Baumberger2009}. The time-axis is scaled, in a sort of coordinate transformation wherein the passage of time is either ``stretched'' or ``compressed'', using a shift factor $\theta$, where $\theta \propto t_{\rm age}^\mu$. This form of shifting works well for non-ergodic systems. It is important to note that $\mu$ and therefore $\theta$ are functions of temperature, and successful shifting can be conducted for data at different temperatures (Fig.~\ref{fig:G-relaxation-temp-time-shifted}).

\begin{figure}[!ht]
\centering
\begin{subfigure}[b]{0.4\textwidth}
	\centering
	(a)
	\includegraphics[width=\textwidth]{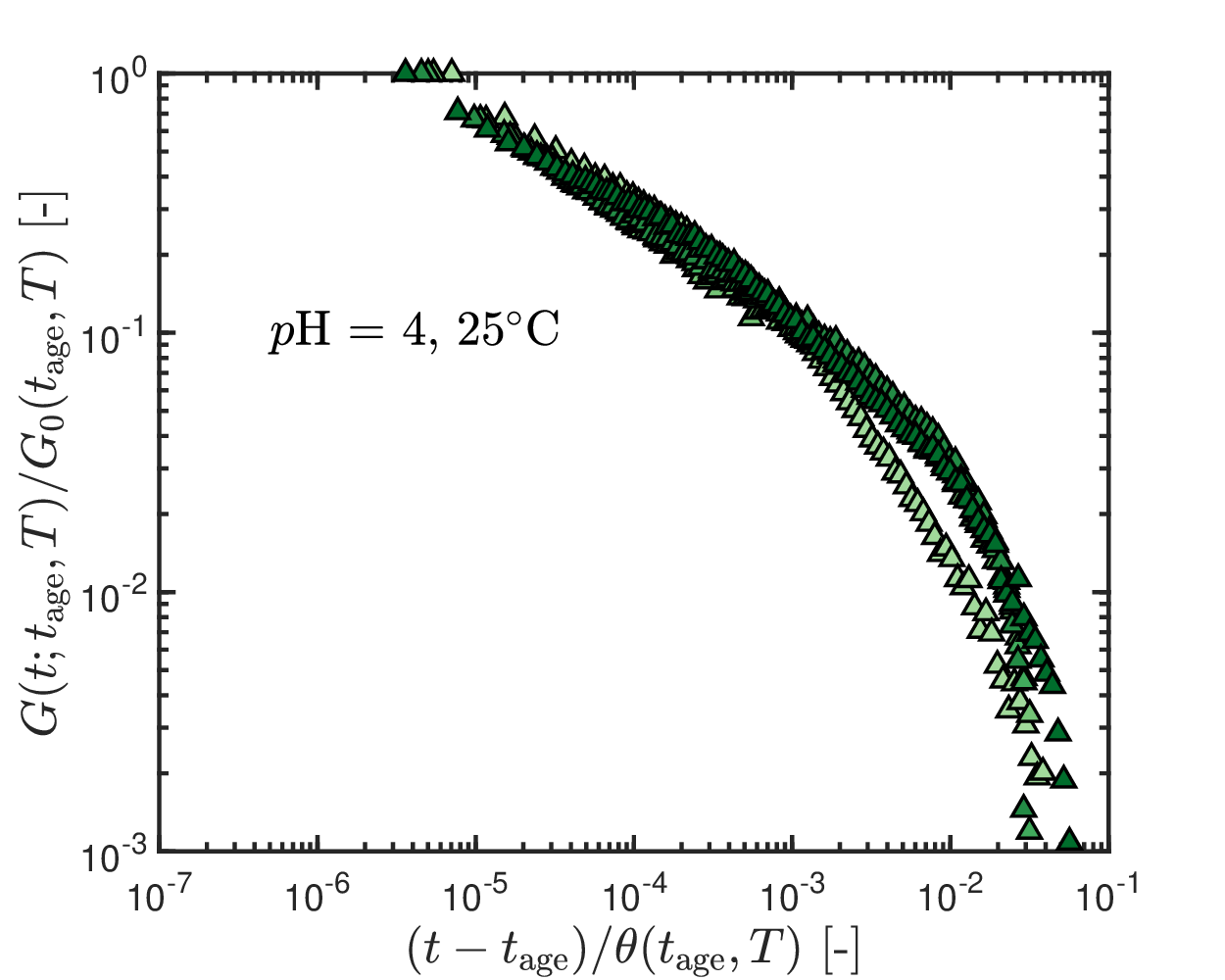}
\end{subfigure}
\begin{subfigure}[b]{0.4\textwidth}
	\centering
	(b)
	\includegraphics[width=\textwidth]{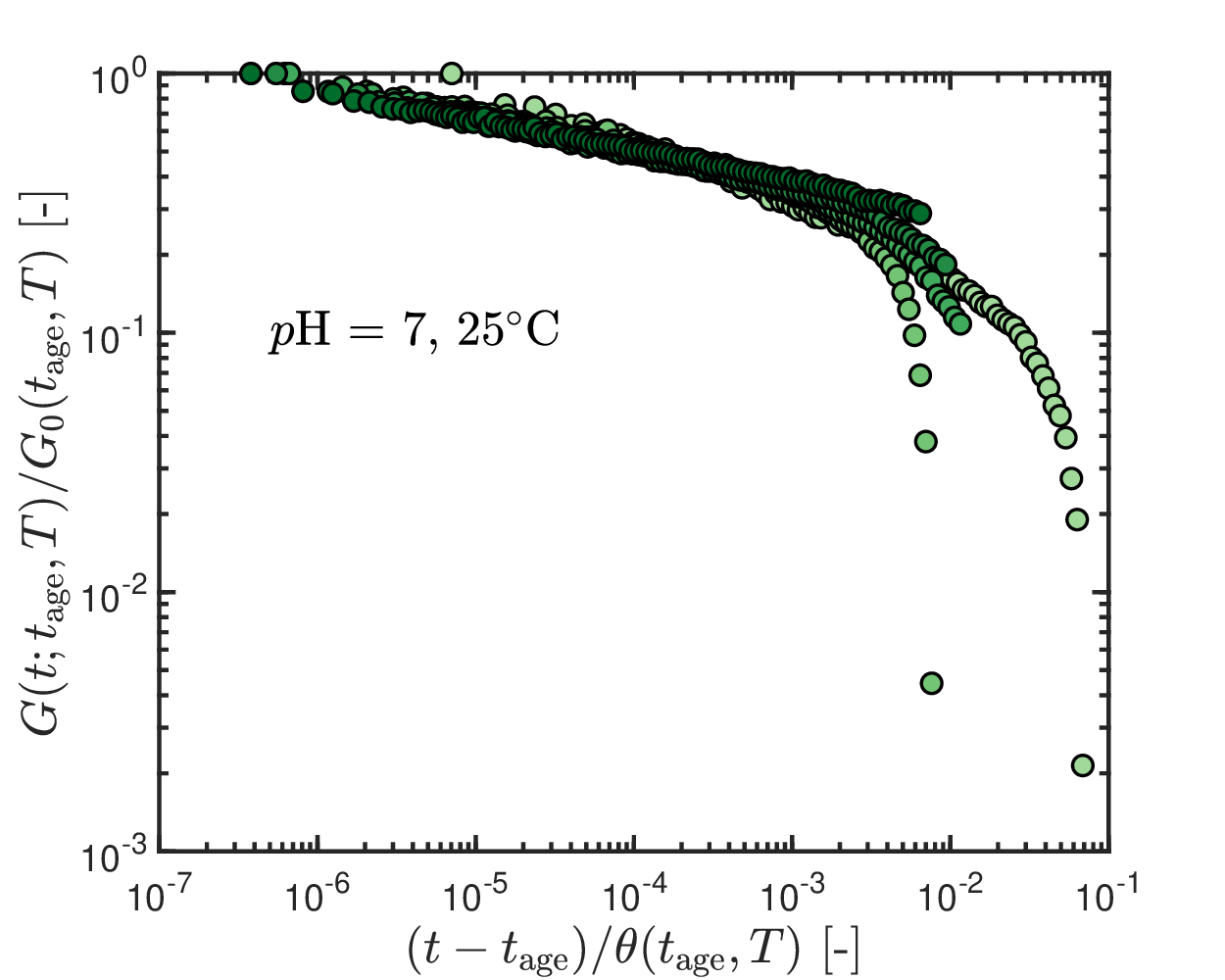}
\end{subfigure}
\caption{Master curves for stress relaxation for gels at different aging times at (a) $p{\rm H} = 4$, and (b) $p{\rm H} = 7$. Such shifting captures non-ergodic effects in the material.}
\label{fig:G-relaxation-pH-time-shifted}
\end{figure}

For stress relaxation data at each temperature, applying the transformation $t_{\rm age} :\rightarrow t_{\rm age}/\theta$ using the appropriate value of $\mu$ in $\theta \propto t_{\rm age}^\mu$ collapses the data onto master curves ({\em n.b.}, the modulus data has also been normalized by its plateau value to achieve this). The same has been repeated with success for the data at two different $p$H values (Fig.~\ref{fig:G-relaxation-pH-time-shifted}).

\begin{figure}[!ht]
\centering
\begin{subfigure}[b]{0.49\textwidth}
	\centering
	(a)
	\includegraphics[width=\textwidth]{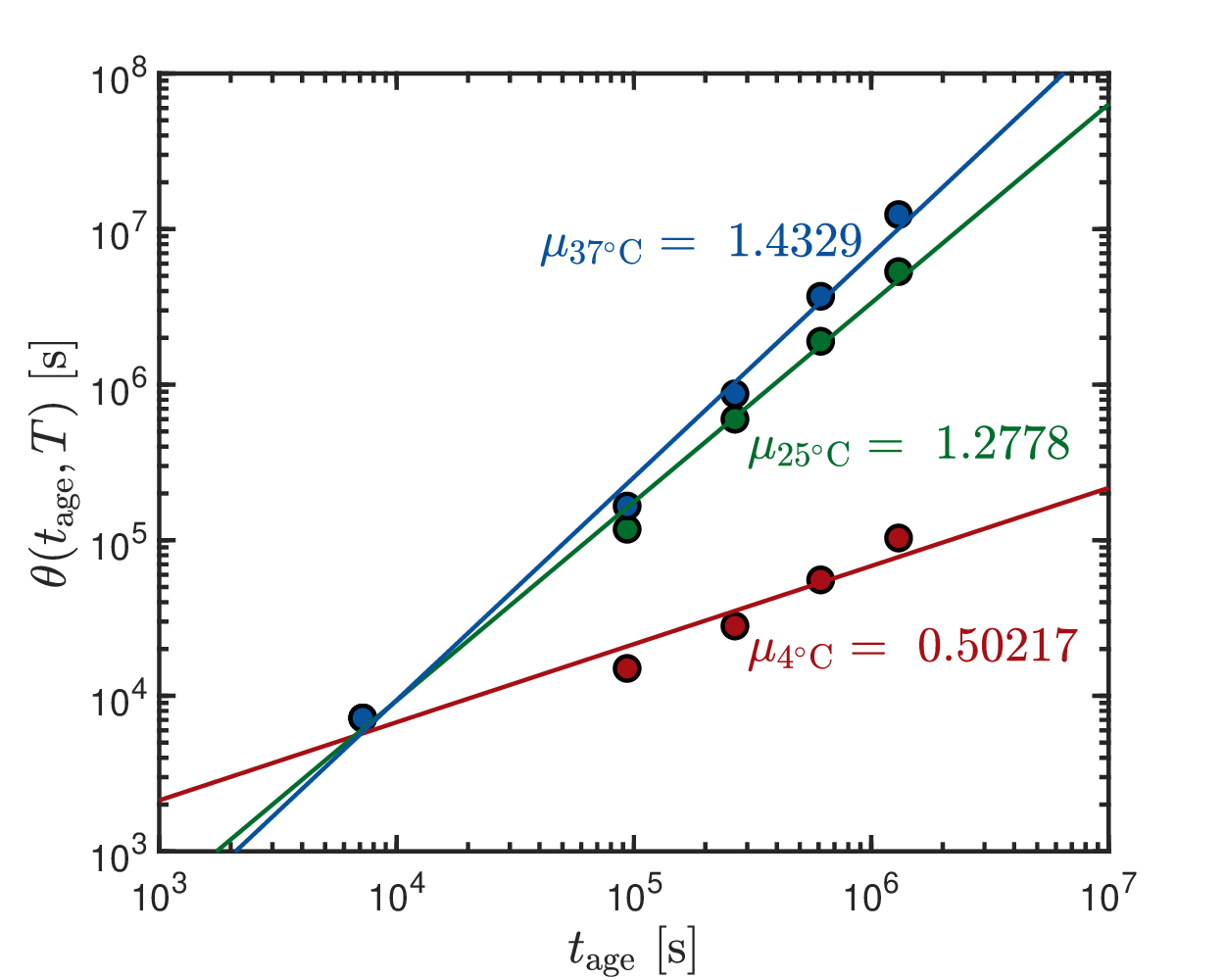}
\end{subfigure}
\hfill
\begin{subfigure}[b]{0.49\textwidth}
	\centering
	(b)
	\includegraphics[width=\textwidth]{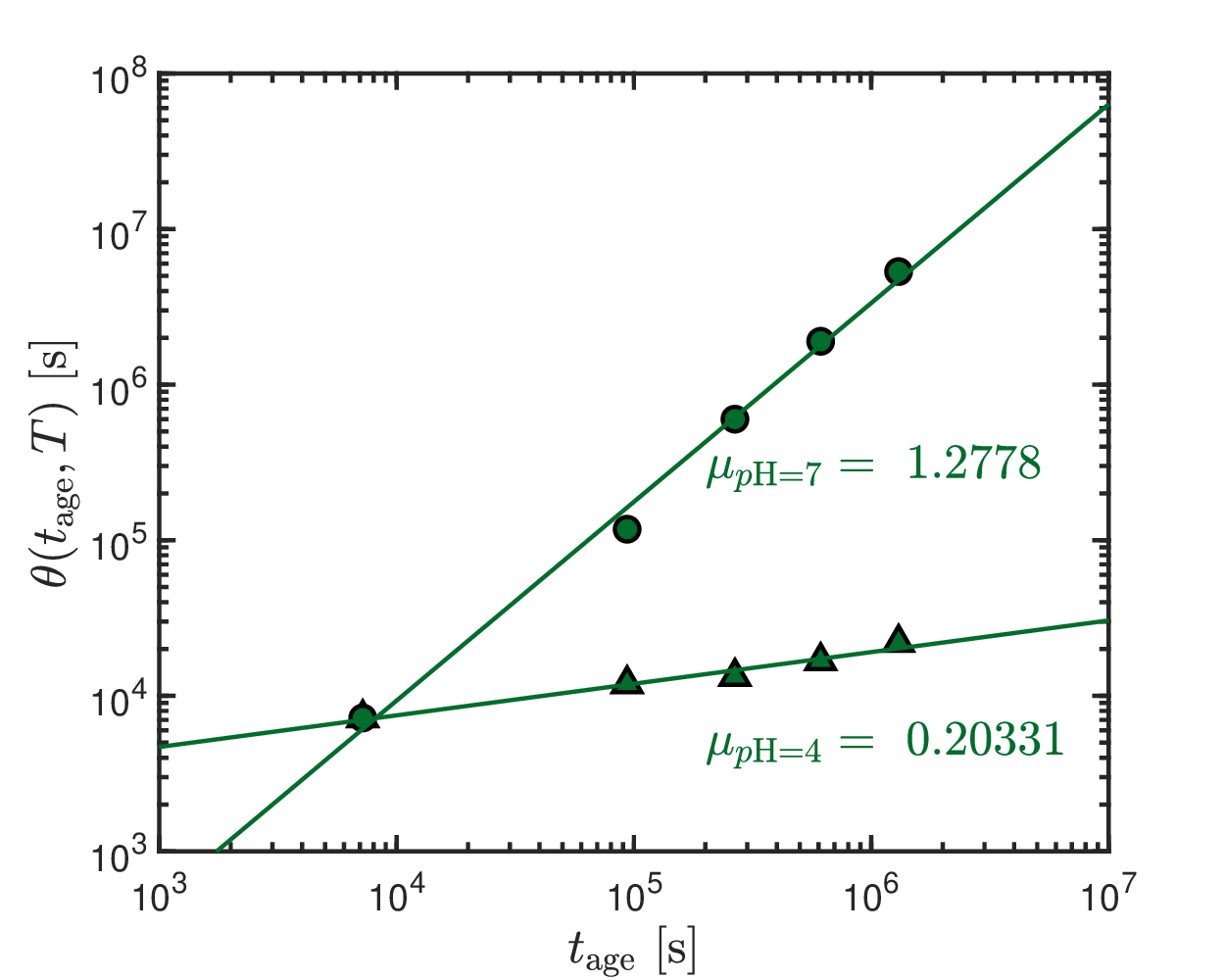}
\end{subfigure}
\caption{Power law correlation between shift factor and aging time, $\theta(t_{\rm age},T) \propto t_{\rm age}^{\mu(T)}$. (a) Extent of correlation between gels at different temperatures, with higher power law slopes at higher temperatures. (b) Extent of correlation between gels at different $p$H, with higher power law slope at higher $p$H.}
\label{fig:G-relaxation-G0-mu}
\end{figure}

The successful shifting is characteristic of non-ergodic systems, and is displayed by systems that age through the growth of interconnected structure elements, in this case being crosslinks between the cellulose chains facilitated at the surface of the CSPs. Such aging systems also tend to be strongly enthalpic (and therefore strongly dependent on temperature), with entropy contributing much less to the free energy. 

The correlation between $\theta$ and $t_{\rm age}$ is a power law (Fig.\ref{fig:G-relaxation-G0-mu}), and we see that $\mu$ is higher at higher $T$, which suggests a stronger correlation and therefore stronger aging. These observations are in full agreement with the rheological data discussed earlier. The same trend is maintained for $p$H dependence, where $\mu$ is higher at higher $p$H, again suggesting a stronger correlation and therefore stronger aging. To further support these observations, we sought to gain structural insight into the effect of aging on the network characteristics by studying the diffusion of polymeric cargo through the gel, as discussed next.


\subsection{\label{subsec:mesh_size}Structural insights into hydrogel aging from diffusivity}
\begin{figure}[ht]
\centering
\includegraphics[width=0.9\textwidth]{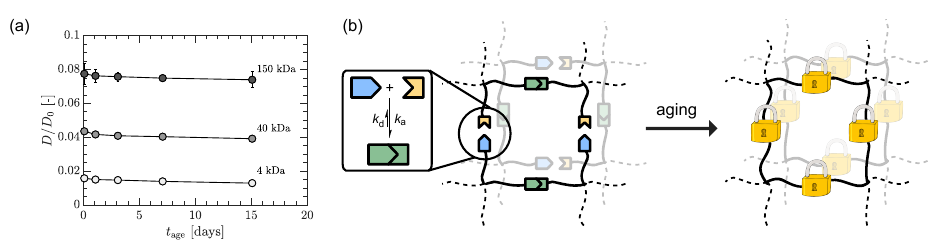}
\caption{(a) Diffusivity data for three cargo sizes obtained from FRAP experiments as functions of aging time. (b) Schematic for the hypothesized dynamic-to-covalent transition scheme for the hydrogel mesh, leading to aging.}
\label{fig:FRAP}
\end{figure}

The observed trends in rheology under the effect of temperature and $p$H is suggestive of the nature of interfacial reactions underlying the aging of the hydrogel. Accordingly, we hypothesized that the number density of crosslinks between polymer chains at particle surfaces would remain largely the same upon mixing the components to make the hydrogel. Aging would not form new crosslink junctions, but only convert a fraction of dynamic noncovalently bound sites into covalently bound anchors. Since the average mesh size of the polymer matrix is determined by the number density of crosslinks, we hypothesized this system would maintain a rather consistent mesh size, and the stiffening only stems from an aging of each bond rather than an increase in the number of bonds. This process is illustrated schematically in Fig.~\ref{fig:FRAP}. Note that immediately upon mixing, transient multivalent crosslinks form within the polymer network \emph{via} chain adsorption at the particle surfaces. The particle surfaces, therefore, are already populated with polymer chains and the number density of polymer-particle interactions has already reached its maximum extent. Yet, while these polymer-particle interactions are dynamic and non-covalent, they can act as anchoring sites. In this paper, when we refer to crosslinks, we shall mean network anchor points which irreversibly connect chains \emph{via} covalent bonding with particle surfaces. It must be kept in mind that, on average, the total number of network anchor points remains the same, yet the fraction of covalent crosslinks increases while the fraction of dynamic bonds decreases as the material ages as polymer-particle interactions undergo a dynamic-to-covalent transition.

To evaluate our hypothesis, we used diffusivity measurements of polymeric solutes through the hydrogel using fluorescence recovery after photobleaching (FRAP) tests (Fig.~\ref{fig:FRAP}a). For FRAP measurements, we used dextran probes conjugated with a fluorescin-isothiocyanate dye (FITC-dextran) as the solute which diffuses through the hydrogel mesh. Three average molecular weights of the polymer were used (4, 40, and 150~kDa), which correspond to three solute ``sizes'' pertaining to their radii of gyration $R_{\rm g}$ in solution (1.4, 4.5, and 8.8~nm respectively). The diffusion coefficient $D$ as a function of aging time $t_{\rm age}$ of each solute obtained from FRAP is compared against the ideal Stokes-Einstein diffusivity, $D_0 = k_{\rm B}T / 6 \pi \eta_{\rm s} R_{\rm g}$, where $\eta_{\rm s}$ is the solvent viscosity, taken as 1~mPa$\cdot$s for water, $k_{\rm B}$ is the Boltzmann constant, and $T$ is the absolute temperature of the system (25$^\circ$C). We see that $D/D_0$ negligibly changes over time (i.e. $<15\%$ change in diffusivity over 15~days) for all three solute sizes. In contrast, rheological measurements showed that the shear modulus increased by more than 250\% in the same duration. Since $D/D_0$ of passively diffusing solutes like polymer globules is a function of mesh size \cite{Axpe}, these observations suggest that the average mesh size of the hydrogel matrix does not meaningfully change during aging while the modulus increases significantly by virtue of strengthening of the crosslinking interactions. Below we propose a reaction scheme underlying the slow transition from non-covalent reversible polymer-particle binding in the hydrogels to irreversibly covalent bonded gels while preserving the number density of crosslinks, which is synergistic with the rheology and diffusion data.


\subsection{\label{sec:aging-ftir}Spectroscopic characterization of aging within HEC-MC/CSP gels}
\begin{figure}[!ht]
\centering
\includegraphics[width=0.9\textwidth]{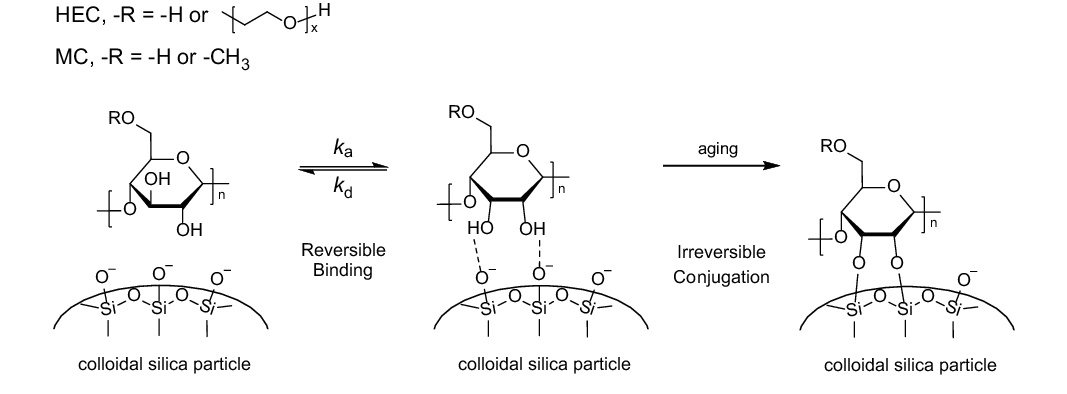}
\caption{Hypothesized interfacial interactions between HEC-MC chains and CSP surface. Shortly after mixing, the polymer chains interact reversibly with the particle surface, which includes hydrogen bonding effects. This slowly transitions to irreversible conjugation as the material ages, the rate of which is dependent on temperature and $p$H.}
\label{fig:structure-aging-pH7}
\end{figure}

We hypothesized that each dynamic, noncovalent bond between the cellulosic hydroxyl groups and the silanol groups on the surface of the CSPs is slowly converted to a \ch{(particle)-Si-O-(polymer)} bond with age, and consequently the topology of the crosslinked chains does not change (Fig.~\ref{fig:structure-aging-pH7}).
In this scheme, the total number of polymer-particle interactions does not change as the hydrogels age, but these interactions only strengthen with time as they undergo the proposed dynamic-to-covalent transition. Such aging has been observed previously in the literature for physical gels of gelatin in water, where chains form a noncovalent network upon dissolution and these junctions undergo a perpetual aging process that results in gel stiffening \cite{Baumberger2009,Normand2000}. Consistent with our hypothesis, the number density of crosslinks within these gelatin-based systems does not change significantly with time, thus preserving the effective mesh size of the network. We then sought to gather direction spectroscopic evidence of the chemical changes occurring during aging and discuss how it supports the proposed reaction scheme and mechanism hypothesized from rheological data described above.

In this section, we monitor the aging processes in our gels with FTIR spectroscopy to generate further insight into the dynamic-to-covalent crosslinking process and quantify the rate of structural stiffening. We focus on the wavenumber range 3200-3600~cm$^{-1}$, since the absorption peak $(a_{\rm m})$ of the cellulosic hydroxyl groups occurs at approximately $\bar{\nu} \approx 3410~{\rm cm}^{-1}$. Per our hypothesis, the condensation reaction between the alcohol groups on the cellulose polymers and the silanol group on the CSP surface would lead to the disappearance of a fraction of these \ch{-OH} bonds and commensurate formation of \ch{(particle)-Si-O-(polymer)} bonds. Consequently, we would expect to observe a decay in this $3410~{\rm cm}^{-1}$ absorption peak as the sample ages. We looked for this signature in aging gels to study both the effects of temperature and $p$H. We only show absorbance data for the relevant range of wavenumbers here, but transmittance data over the full spectral range accessed during our FTIR tests are shown in the SI (Figs.~S3 and S4).

\begin{figure}[!ht]
\centering
\includegraphics[width=0.8\textwidth]{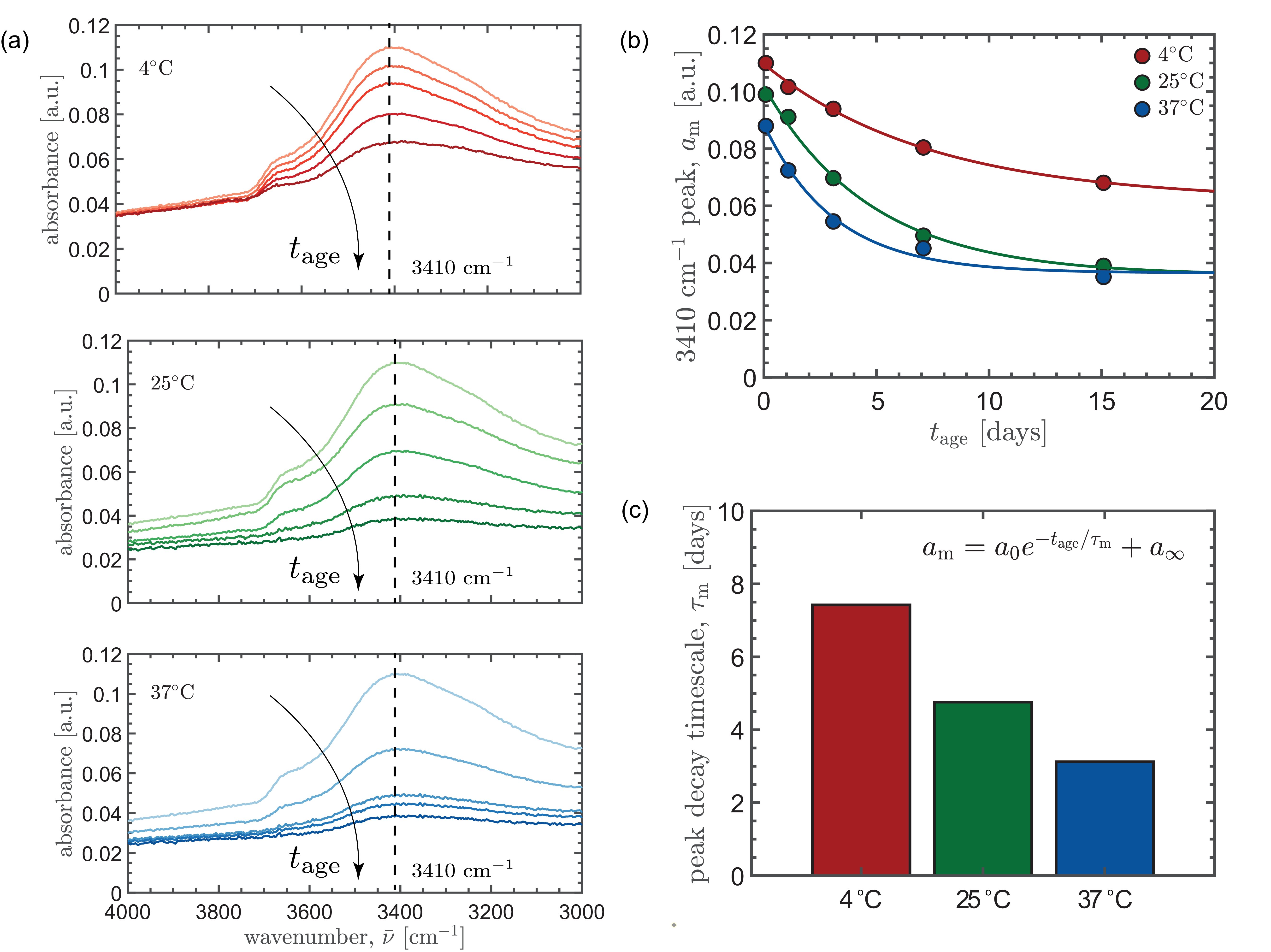}
\caption{FTIR data for gels through aging at different temperatures. (a) Absorbance data for gels at three different temperatures, where aging leads to a decay in the absorbance peak at 3410~cm$^{-1}$, $a_{\rm m}$, which is accelerated as the temperature increases. (b) Decay behavior of $a_{\rm m}$ as function of $t_{\rm age}$ and temperature, showing faster disappearance of the cellulosic hydroxyl peak at higher temperatures. Fit lines are for an exponential function. (c) Peak decay timescale $\tau_{\rm m}$ from fitting of the FTIR data to an exponential decay function, showing that the timescale of decay shortens at elevated temperatures.}
\label{fig:FTIR-gels-abs}
\end{figure}

\begin{figure}[!ht]
\centering
\includegraphics[width=\textwidth]{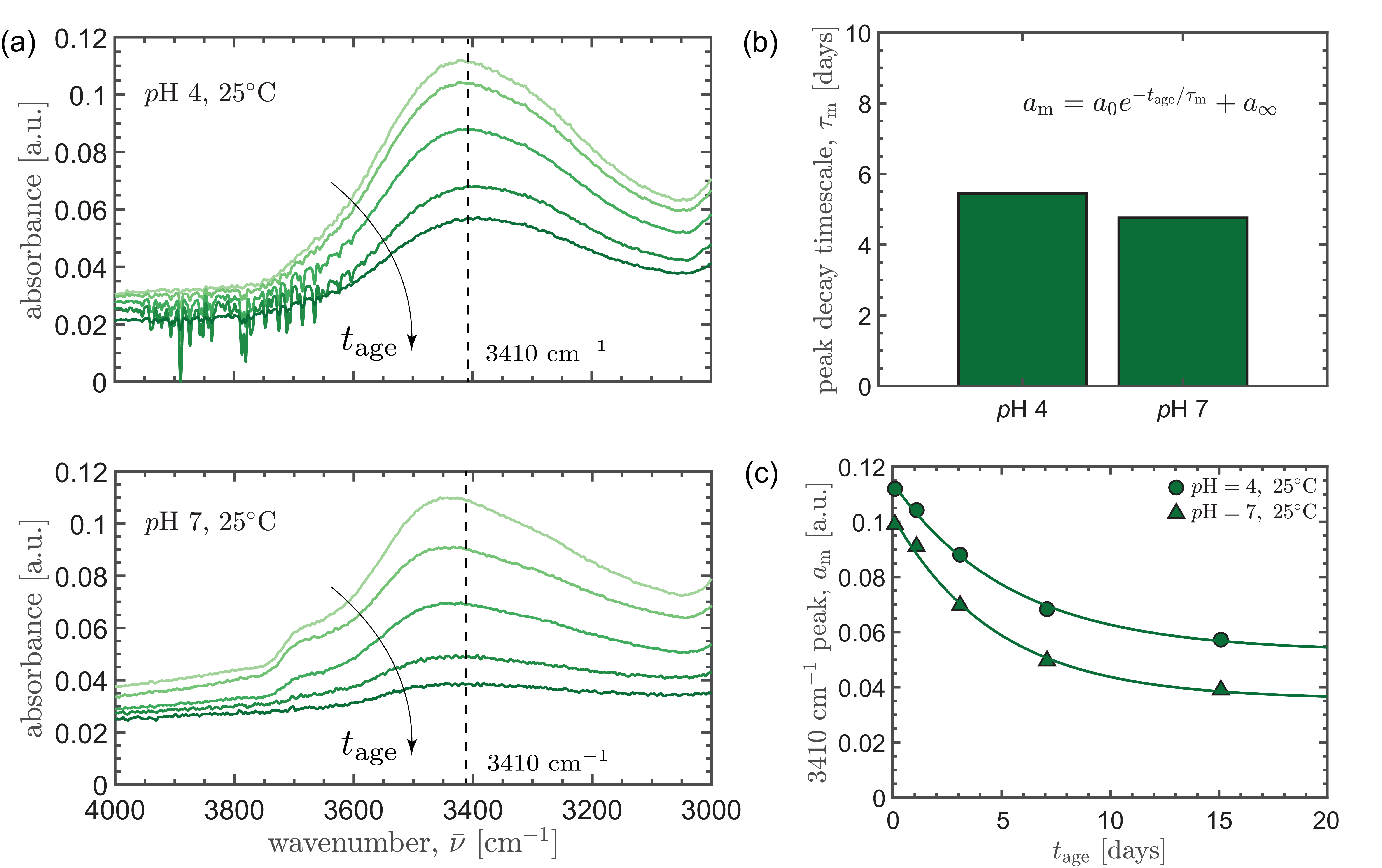}
\caption{FTIR data for gels through aging at different $p$H  values. (a) Absorbance data for gels at $p$H of 4 and 7, where aging leads to a decay in the absorbance peak at 3410~cm$^{-1}$, $a_{\rm m}$, which is accelerated as the $p$H of the gel increases. (b) Decay behavior of $a_{\rm m}$ as function of $t_{\rm age}$ and $p$H, showing slower disappearance of the cellulosic hydroxyl peak in more acidic conditions. Fit lines are for an exponential function. (c) Peak decay timescale $\tau_{\rm m}$ from fitting FTIR data to an exponential decay function, showing that the timescale of decay slows in more acidic conditions.}
\label{fig:FTIR-gels-abs-pH}
\end{figure}

For a given temperature, we see that the absorption spectrum flattens with aging time (Fig.~\ref{fig:FTIR-gels-abs}a), and the absorbance value of peak at $3410~{\rm cm}^{-1}$, $a_{\rm m}$, diminishes with $t_{\rm age}$. As the temperature at which the gels are stored is increased from 4 to 25 to 37$^\circ$C, the peak diminishes more rapidly (Fig.~\ref{fig:FTIR-gels-abs}b). This decrease in the peak absorbance can be captured by a simple exponential decay. The decay timescale, $\tau_{\rm m}$, clearly decreases as the temperature of gel aging increases, indicating faster aging at elevated temperatures (Fig.~\ref{fig:FTIR-gels-abs}c). These observations are in direct agreement with the rheology and kinetics modeling data described above, which showed an increase in stiffness with age resulting from a proliferation of covalent crosslinks over time.

Similarly, as the $p$H of the gels is increased from 4 to 7, the absorbance spectrum flattens out more rapidly (Fig.~\ref{fig:FTIR-gels-abs-pH}a), and the peak of the spectrum, $a_{\rm m}$, decays more quickly (Fig.~\ref{fig:FTIR-gels-abs-pH}b). This decrease in the peak absorbance has been captured by a simple exponential decay, and the decay timescale also decreases with increased $p$H, thus showing an accelerated decay and therefore faster aging in more alkaline conditions (Fig.~\ref{fig:FTIR-gels-abs-pH}c). These FTIR spectroscopy studies support the proposed reaction scheme and enable quantification of the rate and extent of dynamic-to-covalent transitions occurring over time for various aging conditions. Finally, for further quantification, we model the kinetics of this proposed reaction, and infer the progression of crosslink formation during aging through the rheological data shown here. We relate the chemical crosslinking strength to plateau modulus of the gels, giving insight into the energetics of the process.


\subsection{\label{sec:modeling}Modeling of aging processes}
We propose a simple modeling scheme for the chemical reaction underlying the observed aging behaviors to quantify the process. The reaction kinetics associated with the crosslink proliferation is connected back to rheological signatures using network theory, which allows us to estimate energetics of the covalent crosslinking process.

The evolution of the hydrogel network can be modeled using a second-order reaction kinetics model. For macromolecular crosslinking systems involving two reactants, the gelation kinetics have been compared to the predictions of a simple second-order reaction kinetics scheme in the literature \cite{Normand2000,Zimmerman,XuZimmerman}. In our HEC-MC/CSP gels, where the cellulosic polymers and CSP form the two components forming crosslinks by reacting according to the reaction given by
\begin{align}\label{eq:2order-rxn}
\mathcal{A} + \mathcal{B} &\xrightleftharpoons[k_{\rm d}]{k_{\rm a}} \mathcal{C},
\end{align}
with $\mathcal{A}$ and $\mathcal{B}$ representing the two reactants (cellulose polymers and colloidal silica particles), and $\mathcal{C}$ represents the dynamically crosslinked gel product, with $k_{\rm a}$ and $k_{\rm d}$ the rate constants for crosslink association and dissociation respectively. While the gel would be expected to have a relatively small dissociation rate constant compared to the association rate constant, we assume it to be non-zero for the most general case. The differential equation for the rate kinetics appropriate to this scheme is given by
\begin{align}\label{eq:2order-rxn-ode}
\dfrac{{\rm d}}{{\rm d}t} \xi &= k_{\rm a}(a_0 - \xi)(b_0 - \xi) - k_{\rm d} \xi,
\end{align}
where $a_0$ and $b_0$ are the reactive site concentrations of $\mathcal{A}$ and $\mathcal{B}$ respectively, and $\xi$ is some measure of the degree of crosslinking (e.g., the volume density of crosslinks, often called the crosslink concentration). Eq.~\ref{eq:2order-rxn-ode} can be rewritten as
\begin{align}\label{eq:2order-rxn-ode_alt}
\dfrac{{\rm d}}{{\rm d}t} \xi &= k_{\rm a}(a - \xi)(b - \xi),
\end{align}
and the solution to this differential equation is
\begin{align}\label{eq:2order-rxn-ode_alt-soln}
\xi &= a \dfrac{1 - \exp\left[k_{\rm a}(a-b)t\right]}{1 - \dfrac{a}{b} \exp\left[k_{\rm a}(a-b)t\right]},
\end{align}
which can be rewritten as
\begin{align}\label{eq:2order-rxn-ode_alt-soln2}
\xi &= a \dfrac{1 - \exp \left(\kappa t\right)}{1 - \rho \exp \left(\kappa t\right)},
\end{align}
where $\kappa = k_{\rm a}(a-b)$ and $\rho = a/b$. Here, $\kappa$ is a rescaled rate of reaction, and $\rho$ is the molar ratio of the two reactants. Using $\kappa$, which is expected to be a function of temperature, we can capture the rate of crosslink formation and therefore the speed and extent of aging in the gels, and thus connect it to the shear modulus of the material using network theory ({\em vide infra}).


\subsection{\label{subsec:modeling-rheology}Relating kinetics to rheology using network theory}
\begin{figure}[!ht]
\centering
\begin{subfigure}[b]{0.49\textwidth}
	\centering
	(a)
	\includegraphics[width=\textwidth]{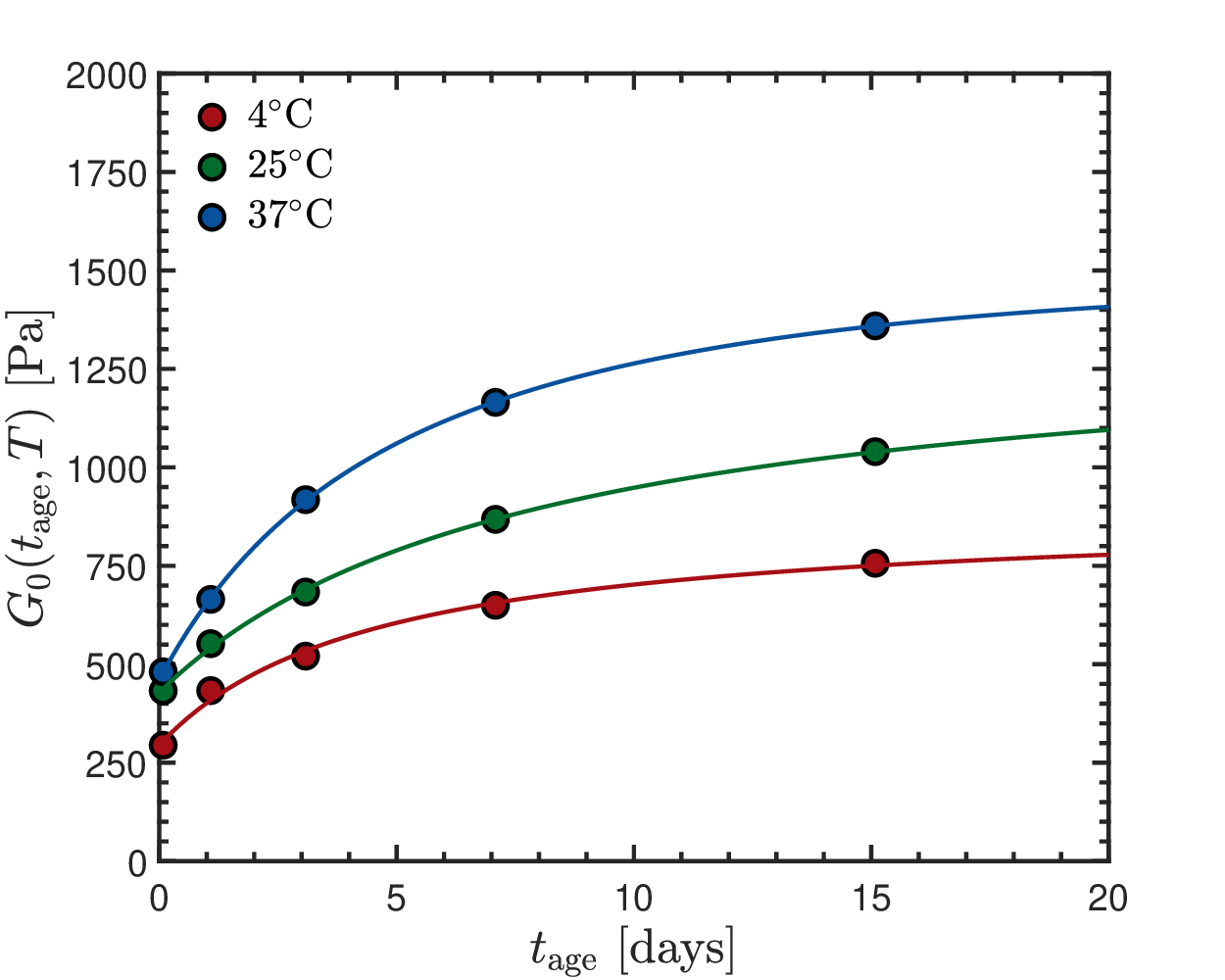}
\end{subfigure}
\hfill
\begin{subfigure}[b]{0.49\textwidth}
	\centering
	(b)
	\includegraphics[width=\textwidth]{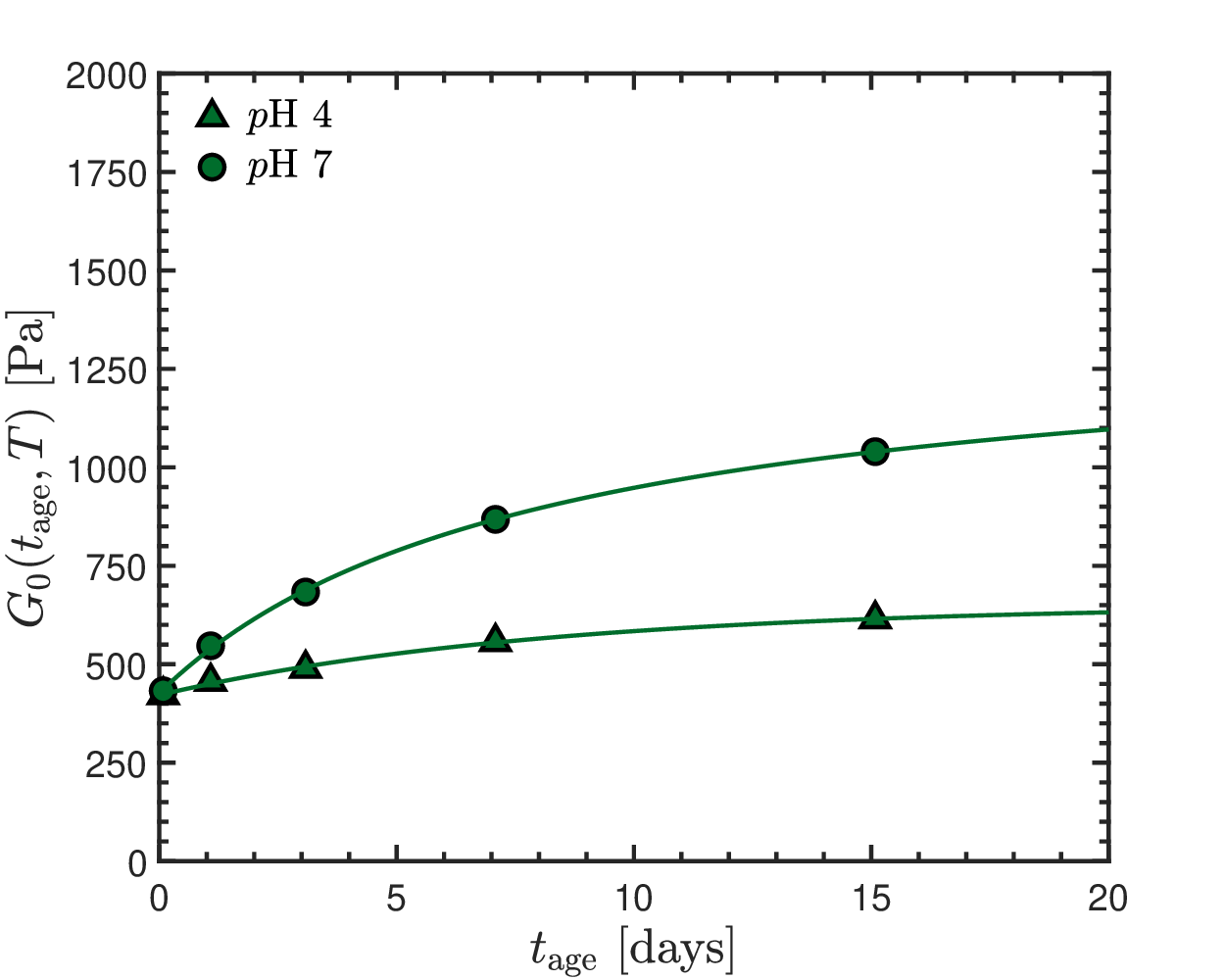}
\end{subfigure}
\caption{(a) Effect of temperature on aging. Plateau modulus as a function of aging time for both temperatures, showing that ages progresses slower at a lower temperature. (b) Effect of $p$H on aging. Plateau modulus as a function of aging time for both temperatures, showing that ages progresses slower at a lower $p$H. Also shown are fits to the second order reaction model, Eq.~\ref{eq:G0-fit}.}
\label{fig:G-relaxation-G0}
\end{figure}

From the stress relaxation data, the plateau modulus $G_0$ is defined as
\begin{align}
G_0(t_{\rm age}, T) \equiv \lim_{t \rightarrow 0} G(t; t_{\rm age}, T),
\end{align}
and it is a function of aging time $t_{\rm age}$ and dependent on temperature $T$. From plots of $G_0$ at different temperatures (Fig.~\ref{fig:G-relaxation-G0}a) and $p$H (Fig.~\ref{fig:G-relaxation-G0}b), it is clear that the gels stiffen with aging time, and the stiffening is stronger at higher temperatures and higher $p$H. These observations are in full agreement with the experimental results described above.

This modulus can also be related to the degree of crosslinking using network theory, where the network stiffness is proportional to the degree of crosslinking \cite{Normand2000,Ferry,DPL_vol1,RubinsteinColby,Graessly,Graessly1975,Chompff,Zimmerman,XuZimmerman}. In short, we have
\begin{align}
G_0(t_{\rm age}, T) \propto \xi(t_{\rm age}, T).
\end{align}
Leveraging the second-order model solution for $\xi$ described above, we can rewrite Eq.~\ref{eq:2order-rxn-ode_alt-soln2} into our model for the aging modulus as
\begin{align}\label{eq:G0-fit}
G_0(t_{\rm age}, T) &= G_{\rm b}(T) + G_{\xi}(T) \dfrac{1 - \exp \left[\kappa(T) t_{\rm age}\right]}{1 - \rho \exp \left[\kappa(T) t_{\rm age}\right]},
\end{align}
where $G_{\rm b}$ is the background viscoelastic modulus independent of the crosslinks (from nonreacting and purely viscoelastic components) which does not change with $t_{\rm age}$ but is dependent on temperature, and $G_\xi$ is the crosslink-dependent modulus, which changes with time/age, and is also dependent on temperature. When fitting Eq.~\ref{eq:G0-fit} to the modulus data for our gels (Fig.~\ref{fig:G-relaxation-G0}), we see that the simple second order reaction kinetics model captures the aging modulus very well. This approach is useful since the fit parameter $\kappa$ is common for the modulus and the crosslink density equations, which means we can use the fit to rheology data to infer information about crosslinking reaction energetics.


\subsection{\label{subsec:modeling-Ea}Activation energy of aging reaction}
\begin{figure}[!h]
\centering
\includegraphics[width=0.5\textwidth]{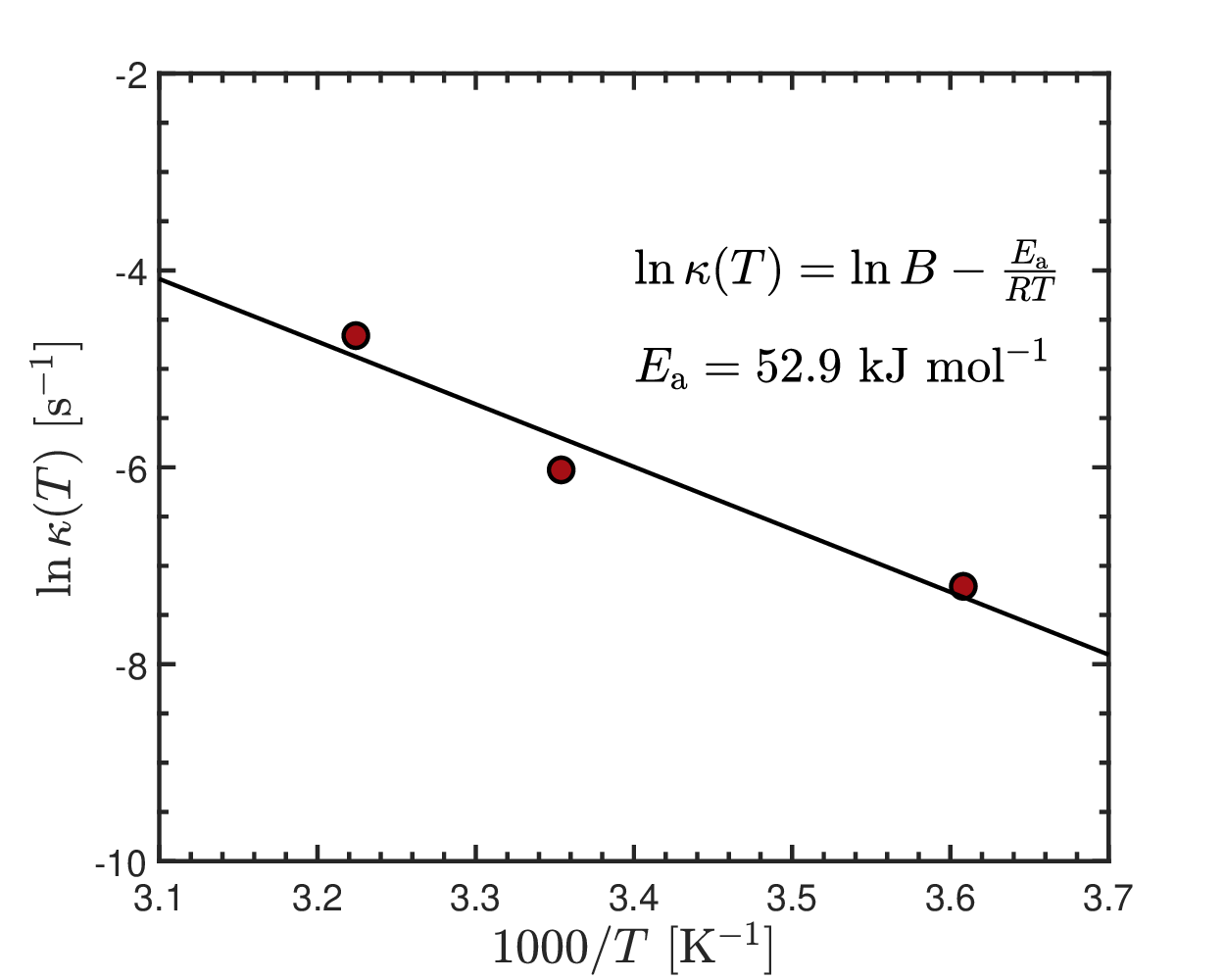}
\caption{Activation energy of the irreversibly conjugating HEC-MC/CSP hydrogel reaction, obtained by connecting the rheological fit parameter $\kappa$ using the second order model to the Arrhenius dependence of equilibrium rate constant.}
\label{fig:Ea-comparison}
\end{figure}

From the solution of the second order reaction (Eq.~\ref{eq:2order-rxn-ode_alt-soln2}), we had defined
\begin{align}
\kappa = k_{\rm a}(a-b),
\end{align}
and also, by definition, the equilibrium rate constant of the reaction is
\begin{align}
K_{\rm eq} \equiv \frac{k_{\rm a}}{k_{\rm d}},
\end{align}
which gives the relation of the equilibrium rate constant and the rate of reaction to be
\begin{align}
\kappa = K_{\rm eq} k_{\rm d}(a-b).
\end{align}
Assuming an Arrhenius rate activated process, we have
\begin{align}
K_{\rm eq}(T) = A \exp \left( - \frac{E_{\rm a}}{RT} \right),
\end{align}
which finally gives
\begin{align}
\kappa(T) = A k_{\rm d} (a-b) \exp \left( - \frac{E_{\rm a}}{RT} \right) \equiv B \exp \left( - \frac{E_{\rm a}}{RT} \right).
\end{align}
From this, we can write
\begin{align}
\ln \kappa(T) = \ln B - \frac{E_{\rm a}}{RT}.
\end{align}
With this equation we can now calculate the activation energy of the covalent crosslinking reaction driving aging in these systems from the reaction rate parameter $\kappa$ obtained from fitting rheology data to our second order model. We can fit the $\ln \kappa(T)$ vs. $1/T$ data to a linear function, and calculate the activation energy $E_{\rm a}$ from the slope of this line (Fig.~\ref{fig:Ea-comparison}).

From the fit, the activation energy is calculated to be $E_{\rm a} = 52.9~{\rm kJ}~{\rm mol}^{-1}$. Compared to the free energy of dynamic crosslink formation in these HEC-MC/CSP gels \cite{AnthonyPNAS2016}, which was determined to be $\Delta G_0 = -18.9~{\rm kJ}~{\rm mol}^{-1}$, the transition to covalent crosslinking has a much higher activation energy, further supporting that this is a temperature activated process.


\section{\label{sec:discussion}Discussion}
In this work, we developed an entropy-driven physical hydrogel system that undergoes a unique dynamic-to-covalent transition in the crosslinking that mimics behaviors observed in many naturally occurring extracellular matrix proteins. Immediately upon mixing, cellulose derivatives (HEC and MC) and CSPs engage in multivalent, dynamic interactions as the polymers adsorb to the interface of the particles, leading to the formation of HEC-MC/CSP hydrogels. Yet, these polymer-particle interactions undergo condensation reactions leading to covalent bond formation between the cellulose polymer chains and the CSP surface. This transitions drives aging of the materials, whereby they get stiffer over time i.e., the modulus increases with aging time), yet without any observable alteration of the hydrogel structure and mesh size. These chemical reactions occurring only where the dynamic polymer-particle interactions are already formed are accelerated rates at elevated temperatures and higher $p$H values. These underlying chemical changes responsible for aging in the HEC-MC/CSP gels is supported with both rheology and spectroscopy data, suggests a slow proliferation of covalent crosslinking \emph{via} interfacial interactions over time. 

Such unique behavior suggests that the extent of crosslinking \emph{via} irreversibly conjugated bonds and consequently the mechanical strength of this system can thus be tuned in multiple ways. Firstly, different temperatures age the gels differently, and a suitable duty cycle of temperatures can be employed to achieve various rates of aging. Secondly, diminishing the density of hydroxyl groups on the cellulose polymers by using alternative cellulosic variants (e.g., methoxy rather than hydroxyl functionality) should directly inhibit the dynamic-to-covalent transitions and associated aging. One such example is using hydroxyethylmethylcellulose (HEMC) instead of HEC or MC as it has fewer hydroxyl groups along the polymer chain as they are replaced by methoxy groups, which dramatically attenuates aging (Fig.~S1). Thirdly, varying the $p$H can also modify the aging behavior. Therefore, depending on the needs in any given application area, slower aging gels can be subjected to elevated temperatures for speeding up the aging process which is otherwise stunted due to substitution or $p$H.

Using time-aging and temperature-stiffening hydrogels based on covalent crosslinking energetics through engineered interfacial interactions, a complementary paradigm to temperature-softening enthalpy-dominated dynamic hydrogels is thus established. We have engineered a transition from an entropy-dominated dynamic crosslinking system to a covalently crosslinked hydrogel system without altering the structure of the matrix, the strength of which can further be tuned based on chemistry (choice of cellulose modification), formulation ($p$H), and temperature. The biomimetic non-ergodic aging through dynamic-to-covalent transitions in these synthetic physical hydrogels enable the development of materials with structures defined by the self-assembly of the gel components that then evolve into versatile covalent hydrogels with tunable mechanical properties. The properties of these systems can be modulated through minor changes in formulation and storage conditions, making them amenable to numerous applications.

\section*{Acknowledgement}
The rheology and spectroscopy data in this paper was collected at the Stanford Soft and Hybrid Materials Facility (SMF) (supported by the NSF grant National Science Foundation grant ECCS-1542152). We thank the Gordon \& Betty Moore Foundation for their financial support of this work as part of our efforts to develop wildland fire retardants to improve the execution of prescribed burns and enhance forest management.

%

\bibliographystyle{unsrt}
\bibliography{references}

\end{document}